\documentclass[twocolumn]{aastex63} 

\usepackage{graphicx} 
\usepackage{float} 
\usepackage{wrapfig} 
\usepackage{lipsum} 
\usepackage{hyperref}
\usepackage{times}
\usepackage{amsmath}
\usepackage{graphicx}
\usepackage{subfigure}
\usepackage{placeins}
\usepackage{hyperref}
\usepackage{gensymb}
\usepackage{upgreek}
\usepackage{natbib}

\usepackage{graphicx}
\usepackage{subfigure}
\usepackage{multirow}
\usepackage{comment}
\usepackage{natbib}
\usepackage{hyperref}
\usepackage{mathtools}
\usepackage{mathrsfs}
\usepackage{fontenc}
\usepackage{color}
\usepackage{url}
\usepackage{hyperref}
\usepackage{gensymb}
\usepackage{pifont}
\graphicspath{{./}}

\newcommand       \K            {\,{\rm K}}
\newcommand       \mum          {{\rm \mu m}}

\usepackage{afterpage}

\shorttitle{AGN Feedback Revealed by JWST Spectroscopy}
\shortauthors{Zhang et al.}

\begin{document}

\title{GATOS: Distinct Feedback Modes in AGN Central Regions Revealed by Spatially Resolved JWST Spectroscopy}

\author[0000-0003-4937-9077]{Lulu Zhang}
\affiliation{The University of Texas at San Antonio, One UTSA Circle, San Antonio, TX 78249, USA; lulu.zhang@utsa.edu; l.l.zhangastro@gmail.com}

\author[0000-0001-7827-5758]{Chris Packham}
\affiliation{The University of Texas at San Antonio, One UTSA Circle, San Antonio, TX 78249, USA; lulu.zhang@utsa.edu; l.l.zhangastro@gmail.com}
\affiliation{National Astronomical Observatory of Japan, National Institutes of Natural Sciences (NINS), 2-21-1 Osawa, Mitaka, Tokyo 181-8588, Japan}

\author[0000-0002-4457-5733]{Erin K. S. Hicks}
\affiliation{Department of Physics and Astronomy, University of Alaska Anchorage, Anchorage, AK 99508-4664, USA}
\affiliation{The University of Texas at San Antonio, One UTSA Circle, San Antonio, TX 78249, USA; lulu.zhang@utsa.edu; l.l.zhangastro@gmail.com}
\affiliation{Department of Physics, University of Alaska, Fairbanks, Alaska 99775-5920, USA}

\author[0000-0002-9627-5281]{Ismael Garc{\'i}a-Bernete}
\affiliation{Centro de Astrobiolog\'{\i}a (CAB), CSIC-INTA, Camino Bajo del Castillo s/n, E-28692 Villanueva de la Ca\~nada, Madrid, Spain}

\author[0000-0001-6794-2519]{Almudena Alonso-Herrero}
\affiliation{Centro de Astrobiolog\'{\i}a (CAB), CSIC-INTA, Camino Bajo del Castillo s/n, E-28692 Villanueva de la Ca\~nada, Madrid, Spain}

\author[0000-0003-4949-7217]{Ric I. Davies}
\affiliation{Max-Planck-Institut für extraterrestrische Physik, Postfach 1312, D-85741, Garching, Germany}

\author[0000-0003-1810-0889]{Martin J. Ward}
\affiliation{Centre for Extragalactic Astronomy, Durham University, South Road, Durham DH1 3LE, UK}

\author[0009-0007-6992-2555]{Daniel E. Delaney}
\affiliation{Department of Physics, University of Alaska, Fairbanks, Alaska 99775-5920, USA}
\affiliation{Department of Physics and Astronomy, University of Alaska Anchorage, Anchorage, AK 99508-4664, USA}

\author[0000-0001-9452-0813]{Takuma Izumi}
\affiliation{National Astronomical Observatory of Japan, National Institutes of Natural Sciences (NINS), 2-21-1 Osawa, Mitaka, Tokyo 181-8588, Japan}
\affiliation{Department of Astronomy, School of Science, Graduate University for Advanced Studies (SOKENDAI), Mitaka, Tokyo 181-8588, Japan}

\author[0000-0001-6854-7545]{Dimitra Rigopoulou}
\affiliation{Department of Physics, University of Oxford, Keble Road, Oxford OX1 3RH, UK}
\affiliation{School of Sciences, European University Cyprus, Diogenes street, Engomi, 1516 Nicosia, Cyprus}

\author[0000-0001-8353-649X]{Cristina Ramos Almeida}
\affiliation{Instituto de Astrof{\'i}sica de Canarias, Calle V{\'i}a L{\'a}ctea, s/n, E-38205, La Laguna, Tenerife, Spain}
\affiliation{Departamento de Astrof{\'i}sica, Universidad de La Laguna, E-38206, La Laguna, Tenerife, Spain}

\author[0000-0002-2356-8358]{Omaira Gonz{\'a}lez-Mart{\'i}n}
\affiliation{Instituto de Radioastronom{\'i}a and Astrof{\'i}sica (IRyA-UNAM), 3-72 (Xangari), 8701, Morelia, Mexico}

\author[0000-0003-0483-3723]{Rogemar A. Riffel}
\affiliation{Departamento de F{\'i}sica, CCNE, Universidade Federal de Santa Maria, Av. Roraima 1000, 97105-900, Santa Maria, RS, Brazil}

\author[0000-0001-5231-2645]{Claudio Ricci}
\affiliation{Department of Astronomy, University of Geneva, ch. d'Ecogia 16, 1290, Versoix, Switzerland}
\affiliation{Instituto de Estudios Astrof\'isicos, Facultad de Ingenier\'ia y Ciencias, Universidad Diego Portales, Av. Ej\'ercito Libertador 441, Santiago, Chile}

\author[0000-0003-2566-2126]{Montserrat Villar-Mart{\'i}n}
\affiliation{Centro de Astrobiolog{\'i}a (CAB), CSIC-INTA, Ctra. de Ajalvir km 4, Torrej{\'o}n de Ardoz, E-28850, Madrid, Spain}

\author[0000-0003-2658-7893]{Francoise Combes}
\affiliation{LUX, Observatoire de Paris, Coll{\`e}ge de France, PSL University, CNRS, Sorbonne University, Paris}

\author[0000-0002-4005-9619]{Miguel Pereira-Santaella}
\affiliation{Instituto de F{\'i}sica Fundamental, CSIC, Calle Serrano 123, 28006 Madrid, Spain}

\author[0000-0002-6353-1111]{Sebastian F. Hoenig}
\affiliation{School of Physics and Astronomy, University of Southampton, Southampton SO17 1BJ, UK}

\author[0000-0002-8651-9879]{Andrew J. Bunker}
\affiliation{Department of Physics, University of Oxford, Keble Road, Oxford OX1 3RH, UK}

\author[0000-0001-9379-4716]{Peter G. Boorman}
\affiliation{Max-Planck-Institut für extraterrestrische Physik, Postfach 1312, D-85741, Garching, Germany}

\author[0000-0001-9791-4228]{Enrica Bellocchi}
\affiliation{Departamento de F\'isica de la Tierra y Astrof\'isica, Fac. de CC. F\'isicas, Universidad Complutense de Madrid, 28040 Madrid, Spain}
\affiliation{Instituto de F\'isica de Part\'iculas y del Cosmos IPARCOS, Fac. CC. F\'isicas, Universidad Complutense de Madrid, 28040 Madrid, Spain}

\author[0000-0003-4209-639X]{Nancy A. Levenson}
\affiliation{Space Telescope Science Institute, 3700 San Martin Drive Baltimore, Maryland 21218, USA}

\author[0000-0003-0444-6897]{Santiago Garc{\'i}a-Burillo}
\affiliation{Observatorio Astron{\'o}mico Nacional (OAN-IGN)-Observatorio de Madrid, Alfonso XII, 3, 28014, Madrid, Spain}

\author[0000-0002-6460-3682]{Fergus R. Donnan}
\affiliation{Department of Astronomy and Astrophysics, University of California, San Diego, La Jolla, CA 92093, USA}

\author[0000-0003-3589-3294]{Anelise Audibert}
\affiliation{Instituto de Astrof{\'i}sica de Canarias, Calle V{\'i}a L{\'a}ctea, s/n, E-38205, La Laguna, Tenerife, Spain}
\affiliation{Departamento de Astrof{\'i}sica, Universidad de La Laguna, E-38206, La Laguna, Tenerife, Spain}

\author[0000-0003-4809-6147]{Lindsay Fuller}
\affiliation{The University of Texas at San Antonio, One UTSA Circle, San Antonio, TX 78249, USA; lulu.zhang@utsa.edu; l.l.zhangastro@gmail.com}

\author[0000-0003-0699-6083]{Tanio D{\'i}az-Santos}
\affiliation{Institute of Astrophysics, Foundation for Research and Technology-Hellas (FORTH), Heraklion 70013, Greece}
\affiliation{School of Sciences, European University Cyprus, Diogenes street, Engomi, 1516 Nicosia, Cyprus}

\author[0000-0002-0001-3587]{David J. Rosario}
\affiliation{School of Mathematics, Statistics and Physics, Newcastle University, Newcastle upon Tyne, NE1 7RU, UK}



\begin{abstract}

This manuscript presents JWST MIRI/MRS observations of the central $r \approx 40-240$ pc regions of five active galactic nuclei (AGN) spanning a wide range of luminosities (log\,$L_{\rm bol}/{\rm erg\,s^{-1}} \approx 39.8$--43.8). Combining multiphase diagnostics from polycyclic aromatic hydrocarbons (PAHs), molecular hydrogen (H$_2$), and ionized gas at spatial scales of $\sim$4--24 pc, this study presents a spatially resolved investigation into the effects of the two distinct AGN feedback modes--radiative and kinetic--on the surrounding medium. The results indicate that these two feedback modes, associated with AGN irradiation and shock processing, respectively, collectively drive the relative suppression of PAH emission in the nuclear regions of the targets studied here. Moreover, the coexistence of these two AGN feedback modes, especially the shock processing associated with either jets or outflows, in the central regions of AGN naturally explains both the bimodal distribution of PAH band ratios observed in the targets studied here and the seemingly disparate results reported in the literature. Although based on a limited sample, these findings provide new insights into calibrating star-formation rates (SFRs) from PAH emission in AGN, and more importantly, lay the groundwork for a practical framework to diagnose and quantify AGN feedback in the JWST era. 

\end{abstract}


\section{Introduction}

The co-evolution of supermassive black holes (SMBHs) and their host galaxies--a central paradigm in modern extragalactic astrophysics (for a review see \citealt{Kormendy&Ho2013})--requires the injection of substantial energy and momentum from active galactic nuclei (AGN)--the observable phases of SMBH accretion--into the surrounding interstellar medium (ISM) (for reviews see \citealt{Silk&Rees1998, Fabian2012}). In observation, this injection--commonly referred to as AGN feedback--manifests in two primary modes, widely termed the “radiative” (or quasar) mode and the “kinetic” (or radio) mode (for a review see \citealt{Alexander&Hickox2012}). 

The radiative mode is typically associated with luminous AGN, whose intense radiation efficiently expels the surrounding gas (e.g., \citealt{DiMatteo.etal.2005, Hopkins.etal.2008}), driving multiphase outflows detected across a broad range of wavelengths, including ionized gas traced by UV-optical-infared emission lines, warm molecular gas in the infrared, and cold molecular gas in millimeter wavelengths (e.g., \citealt{Cicone.etal.2014, Davies.etal.2014, Davies.etal.2024, Fiore.etal.2017, Fluetsch.etal.2019, RamosAlmeida.etal.2022, Esposito.etal.2024, Zhang.etal.2024a, Bellocchi.etal.2026}; and for a review see \citealt{Harrison&RamosAlmeida2024}). In contrast, the kinetic mode is more prevalent in low-luminosity AGN, which have extremely low radiative efficiencies (\citealt{Ho.etal.2003, Ho2009}) and whose central engines are often dominated by jet power (e.g., \citealt{Mezcua&Prieto2014, Fernandez-Ontiveros.etal.2012, Goold.etal.2026}), heating the surrounding gas through shocks (e.g., \citealt{Weinberger.etal.2017}; and for reviews see \citealt{McNamara&Nulsen2007, Ho2008}). Direct evidence of such jet- and shock-driven feedback, including cavities, lobes, and shock fronts in hot gas, have been observed in radio and X-ray bands (e.g., \citealt{McNamara&Nulsen2007, Morganti.etal.2013a}), with jet- and shock-ISM interaction signatures, including disturbed kinematics and enhanced excitation in cold and warm gas, been observed in optical, infrared, and millimeter bands (e.g., \citealt{Morganti.etal.2013b, Nesvadba.etal.2017, Venturi.etal.2021, Audibert.etal.2023, Dasyra.etal.2024, Riffel.etal.2026b}).

These two modes are proposed to reflect different accretion states and environmental couplings, yet both play a fundamental role in heating, redistributing, and transforming the ISM on galactic scales, thereby driving galaxy evolution (for reviews see \citealt{Alexander&Hickox2012, Heckman&Best2014}). A clear understanding of AGN feedback is therefore essential for placing meaningful constraints on models of galaxy formation and evolution. Correspondingly, state-of-the-art cosmological simulations increasingly rely on subgrid prescriptions for AGN-driven energy injection to reproduce key observables, such as the galaxy stellar mass function, star formation histories, and atomic/molecular gas fractions (e.g., \citealt{Crain.etal.2015, Pillepich.etal.2018, Dave.etal.2020}). However, significant uncertainties remain regarding how AGN-released energy couples to different gas phases and the spatial and temporal scales over which feedback operates. In this context, detailed studies of nearby galaxies, through spatially resolved measurements of gas content, kinematics, excitation, and energetics, provide critical laboratories for addressing these uncertainties. Insights from the local studies can then be extrapolated to the high-redshift universe, where direct constraints are more limited, thereby informing our understanding of AGN feedback and its role in galaxy formation and evolution in the early universe.

The advent of JWST (\citealt{Gardner.etal.2023}) opens a new window for probing AGN feedback with unprecedented sensitivity and spectral coverage. Its IFU capabilities in the near- and mid-infrared enable spatially resolved, and joint, access to diagnostic features from multiple ionized, neutral, and molecular gas, as well as dust components, in local galaxies. Key spectral tracers, including fine-structure lines from ionized gas, rotational and ro-vibrational transitions of molecular hydrogen (H$_2$), and broad features from polycyclic aromatic hydrocarbon (PAH) molecules, are particularly sensitive to the heating and excitation processes associated with AGN activity (e.g., \citealt{Davies.etal.2024, Garcia-Bernete.etal.2024b, AlonsoHerrero.etal.2025, Lopez.etal.2025, RamosAlmeida.etal.2025, Goold.etal.2026, Hermosa-Munoz.etal.2026, Zhang.etal.2026}). By capturing these signatures at high spatial and spectral resolution, JWST enables a comprehensive, multiphase characterization of feedback processes (e.g., \citealt{Zhang&Ho2023c, Zhang.etal.2024b, Zhang.etal.2026} and references therein), yielding critical insights into how AGN shape their host galaxies and, in turn, drive galaxy evolution.

\section{Observation and Analysis}\label{sec2}

\subsection{Targets and Observations}\label{sec2.1}

The targets analyzed here comprise three low-ionization nuclear emission-line regions (LINERs) and two Seyfert galaxies, spanning a broad range of AGN luminosities ($L_{\rm bol} \approx 10^{39.8} - 10^{43.8}\,{\rm erg\,s^{-1}}$). These targets, unless otherwise specified in Table~\ref{tabinfo}, are part of the Galaxy Activity, Torus, and Outflow Survey\footnote{\url{https://gatos-astro.com}} (GATOS; e.g., \citealt{AlonsoHerrero.etal.2021, Garcia-Burillo.etal.2021, Garcia-Burillo.etal.2024}). This study presents JWST spectroscopic observations of the two Seyfert galaxies for the first time and performs a spatially resolved analysis of all five targets, including the three LINERs for which only nuclear integrated spectra have previously been analyzed  (\citealt{Zhang.etal.2026}). These targets were selected for JWST programs predominantly based on their nuclear sub-kiloparsec-scale PAH properties, which span two regimes expected to arise from the primary modes of AGN feedback. Additionally, the three LINERs all show evidence of radio cores associated with AGN (\citealt{Nagar.etal.2005, Nemmen.etal.2014}), in contrast to the two Seyfert galaxies, which both exhibit prominent AGN ionization cones (\citealt{Marconi.etal.1994, daSilva.etal.2023}); NGC~7314 also hosts two off-nuclear radio spots (\citealt{Thean.etal.2000}). As such, they provide an ideal, albeit limited, sample for disentangling the roles of different AGN feedback mechanisms through the joint interpretation of PAH features and other infrared diagnostics.

\startlongtable
\setlength{\tabcolsep}{6pt}
\begin{deluxetable*}{cccccccccc}
\tablecolumns{10}
\tablecaption{Basic Properties of the AGN Sample}
\tablehead{
\colhead{Target} & \colhead{Hubble Type} & \colhead{AGN Type} & \colhead{\,\,\,\,\,\,\,$z$\,\,\,\,\,\,\,} & \colhead{\,\,\,\,\,\,\,\,\,$D$\,\,\,\,\,\,\,\,\,} & \colhead{\,\,\,scale\,\,\,} & \colhead{\,\,\,\,\,$cos(i)$\,\,\,\,\,} & \colhead{log $L_{\rm bol}$} & \colhead{log $M_{\rm BH}$} & \colhead{log $\lambda_{\rm Edd}$} \\
\colhead{(-)} & \colhead{(-)} & \colhead{(-)} & \colhead{(-)} & \colhead{(Mpc)} & \colhead{(pc/\arcsec)} & \colhead{(-)} & \colhead{[$\rm erg\,s^{-1}$]} & \colhead{[$\rm M_{\odot}$]} & \colhead{(-)} \\
\colhead{(1)} & \colhead{(2)} & \colhead{(3)} & \colhead{(4)} & \colhead{(5)} & \colhead{(6)} & \colhead{(7)} & \colhead{(8)} & \colhead{(9)} & \colhead{(10)} }
\startdata
NGC~1097$^{\dag}$ & SBb & L/Sy & 0.004240 & 17.2 & 80& 0.68 & 42.0 & 8.1 & $-$4.2\\
NGC~3190 & Sa & L & 0.004370 & 24.5 & 120 & 0.34 & 40.8 & 8.2 & $-$5.5\\
NGC~4736$^{\star}$  & Sab & L/Sy & 0.001027 & 5.1 & 25 & 0.81 & 39.8 & 7.2 & $-$5.5 \\
NGC~7314$^{\ddag}$ & SABbc & Sy & 0.004763 & 16.7 & 80 & 0.34 & 43.5 & 6.2 & $-$0.8 \\
Circinus Galaxy$^{\ast}$ & SAb & Sy & 0.001448 & 4.2 & 20 & 0.42 & 43.8 & 6.2 & $-$0.5 \\
\enddata
\tablecomments{\footnotesize Column (1-5): Target name, Hubble type, AGN type (``L'' for LINER; ``Sy'' for Seyfert), redshift, and redshift-independent distance taken from the NASA/IPAC Extragalactic Database; Column (6): The linear scale corresponding to an angular size of 1 arcsec at the distance of each target. Column (7): Cosine value of the inclination angle on the galactic scale, taken as the ratio of the photometric minor to major axis, cos(i) = b/a, unless otherwise specified (\citealt{Kennicutt.etal.2003, Garcia-Burillo.etal.2021}). Column (8): AGN bolometric luminosity $L_{\rm bol} = 15.8\times L_{\rm (2-10)keV}$, with the nuclear 2--10 keV X-ray luminosity taken from \cite{Ho2009} unless otherwise specified; the values in this table have been adjusted to account for differences between the adopted distances and those reported in the cited references for the X-ray luminosities. Column (9): Black hole mass derived from the $M_{\rm BH}-\sigma_{*}$ relation of \cite{Greene.etal.2020} unless otherwise specified, with $\sigma_{*}$ from \cite{Ho.etal.2009}. Column (10): Eddington ratios $\lambda_{\rm Edd} = L_{\rm bol}/L_{\rm Edd}$, with $L_{\rm Edd} = 1.26\times10^{38}\ (M_{\rm BH}/{\rm M_{\odot}})$. [$\dag$]: $L_{\rm (2-10)keV}$ and $M_{\rm BH}$ are taken from \cite{Cisternas.etal.2013}. [$\ddag$]: $L_{\rm (2-10)keV}$ and $M_{\rm BH}$ are taken from \cite{daSilva.etal.2023}. [$\ast$]: Inclination angle $i$, $L_{\rm (2-10)keV}$, and $M_{\rm BH}$ are taken from \cite{Izumi.etal.2018}.  [$\star$]: This target is part of the GATOS sample selected for JWST GO program 4972 but not for new observations, as it was previously observed under GO program 2016.}
\label{tabinfo}
\end{deluxetable*}

The analysis presented here is based on JWST MIRI/MRS IFU spectroscopy (\citealt{Wells.etal.2015, Wright.etal.2023}) of the nuclear $r \approx 40-240$ pc regions of the five AGN targets, corresponding to the $\sim 4\arcsec \times 4\arcsec$ field of view of the observations. The data were acquired through JWST General Observer (GO) programs 2016 (PI: A. Seth; \citealt{Goold.etal.2026}; NGC~4736), 4225 (PI: T. Izumi; the Circinus Galaxy; hereafter Circinus), 4972 (PI: L. Zhang; \citealt{Zhang.etal.2026}; NGC~1097 and NGC~3190), and 7429 (PI: L. Zhang; NGC~7314). All MIRI/MRS observations adopted a 4-point dither with the FASTR1 readout pattern for the source exposures, along with dedicated background exposures obtained using 1-point (non-dither), 2-point dither, or 4-point dither patterns, depending on the program. The total on-source exposure times range from 388 to 1332 seconds, depending on the brightness of each source. The observations of the targets can be accessed via doi: \href{https://doi.org/10.17909/9bgc-da21}{10.17909/9bgc-da21}.

\subsection{Data Processing and Spectral Extraction}\label{sec2.2}

The data were reduced following \cite{Zhang.etal.2026}, using the JWST Science Calibration Pipeline (v1.18.0; \citealt{Bushouse.etal.2025}), with the context 1364 for the Calibration References Data System. In particular, the ${\tt residual\_fringe}$ correction was activated to correct the fringe residuals (\citealt{Law.etal.2023}), and the default ${\tt master\_bg}$ function was adopted for background subtraction. In addition, the ${\tt firstframe}$ option was enabled during the data reduction of Circinus to mitigate the saturation effect from this bright source, particularly in Channel 3 (ch3; $\lambda \approx 11.6 - 18.0\,\mum$). We note that, although this step eliminates saturation effects in the continuum of Circinus, several bright emission lines in ch3, particularly [Ne~{\small III}]15.56$\mu$m, and in a few spaxels, [Ne~{\small V}]14.32$\mu$m, still suffer from saturation near the nucleus. Moreover, H$_2$ rotational transitions $S(1)$ and $S(2)$ in ch3 are dominated by uncertainties associated with the strong continuum, preventing reliable spaxel-based measurements of these two transitions even in the more extended circumnuclear region of this target. We therefore excluded spaxels in these affected regions from the analysis, except where they were included for illustrative purposes only. 

The JWST MIRI/MRS pipeline ultimately produced twelve spectral data cubes across four channels (i.e., ch$1-4$) for each target. Together, these cubes cover the $4.9 - 27.9\,\mum$ wavelength range with a spectral resolution of $\sim 4000 - 2000$ (\citealt{Argyriou.etal.2023, Pontoppidan.etal.2024}). A key requirement for spatially resolved analysis is a uniform angular resolution across all slices of the data cubes. Accordingly, we convolved the MRS data cubes from the first three channels (i.e., ch$1-3$; $\lambda \approx 4.9$–$18.0\,\mum$) to a common point-spread-function (PSF) with a full width at half maximum (FWHM) angular resolution of $0\farcs7$, corresponding to the largest PSF in ch3. The ch4 cubes were retained at their native angular resolution, as the subsequent analysis focuses on emission features covered by the first three channels. Following PSF matching, the spectral data cubes were reprojected onto a common coordinate frame with a pixel size of $0\farcs2$ (i.e., the native pixel size of ch3) prior to spectral extraction. This pixel size corresponds to physical scales of $\sim 4-24$ pc at the distances of the five targets. Spectra extracted from all sub-bands of a given spaxel are stitched together to produce a complete infrared spectrum for each $0\farcs2 \times 0\farcs2$ spaxel, using the ratios of the median fluxes in the overlapping wavelength regions as scaling factors. These factors are generally close to unity, with typical uncertainties of $\sim$10\%.

\subsection{Spectral Decomposition and Measurements}\label{sec2.3}

We first decomposed the PAH features using a multi-component fitting approach applied to the extracted spectra of each spaxel (see Appendix~\ref{secA0} for examples). As detailed by \cite{Zhang.etal.2026}, the model comprises a series of Drude profiles for individual PAH features, multiple modified blackbodies representing dust components, and a 5000~$\rm K$ blackbody to approximate the stellar continuum, with all emission lines masked during the fitting process. All the model components are subject to foreground screen extinction (i.e., $e^{-\tau_{\lambda}}$) in the fitting. We adopted here the infrared optical depth curve (i.e., $\tau_{\lambda}$) from \citeauthor{Smith.etal.2007} (\citeyear{Smith.etal.2007}; their Equation~4), scaled by the fitted free parameter $\tau_{9.7}$, which represents the optical depth at 9.7 $\mum$. The decomposition was performed using the Bayesian Markov Chain Monte Carlo sampler $\tt emcee$ in the $\tt Python$ environment (\citealt{Foreman-Mackey.etal.2013}), with the median and standard deviation of each parameter’s posterior distribution taken as the final estimate and its corresponding uncertainty.

We then fitted H$_2$ transitions and ionized emission lines in each spaxel individually, using single- and double-Gaussian profiles, respectively, together with a local linear continuum. The fits were optimized via the Levenberg–Marquardt least-squares minimization algorithm. To improve the robustness of the measurements, each emission-line spectrum was perturbed with random noise at the level of its uncertainty, and the fitting was repeated 100 times. The median and standard deviation of these 100 fits were adopted as the final flux and its associated uncertainty for each line. Additionally, all emission-line fluxes were then corrected for dust extinction using the same $e^{-\tau_{\lambda}}$ curve fitted for each spaxel from the multi-component full-spectrum fitting described above.

\section{Spatially Resolved Analysis}\label{sec3}

\subsection{Correlation Between PAH Emission and SF Activity}\label{sec3.1}

Among the molecular, atomic, and ionzied gas phases accessible to JWST spectroscopy, PAH molecules exhibit prominent emission features--particularly around 6.2, 7.7, and 11.3 $\mum$--that are widely calibrated as tracers of star formation rate (SFR; e.g., \citealt{Treyer.etal.2010, Shipley.etal.2016, Maragkoudakis.etal.2018, Xie&Ho2019, Zhang&Ho2023b}). However, the intrinsic PAH emission is often relatively suppressed, even deficit, in the harsh environments surrounding AGN (e.g., \citealt{Smith.etal.2007, ODowd.etal.2009, Diamond-Stanic&Rieke2010, Sales.etal.2010, Garcia-Bernete.etal.2015, Garcia-Bernete.etal.2022a, Garcia-Bernete.etal.2024b, Esparza-Arredondo.etal.2018, Lai.etal.2022, Xie&Ho2022, RamosAlmeida.etal.2023, Ogle.etal.2025}), making it also a diagnostic of AGN feedback (e.g., \citealt{Zhang.etal.2022, Zhang.etal.2024b, Zhang.etal.2026} and references therein). In theory, this intrinsic suppression is primarily attributed to the selective erosion and destruction of PAH molecules by the radiative effects of extreme-UV and X-ray photons (\citealt{Aitken&Roche1985, Voit1992}), as well as the mechanical processing by shocks (\citealt{Micelotta.etal.2010a, Micelotta.etal.2010b}).

\begin{figure*}[!ht]
\center{\includegraphics[width=1\linewidth]{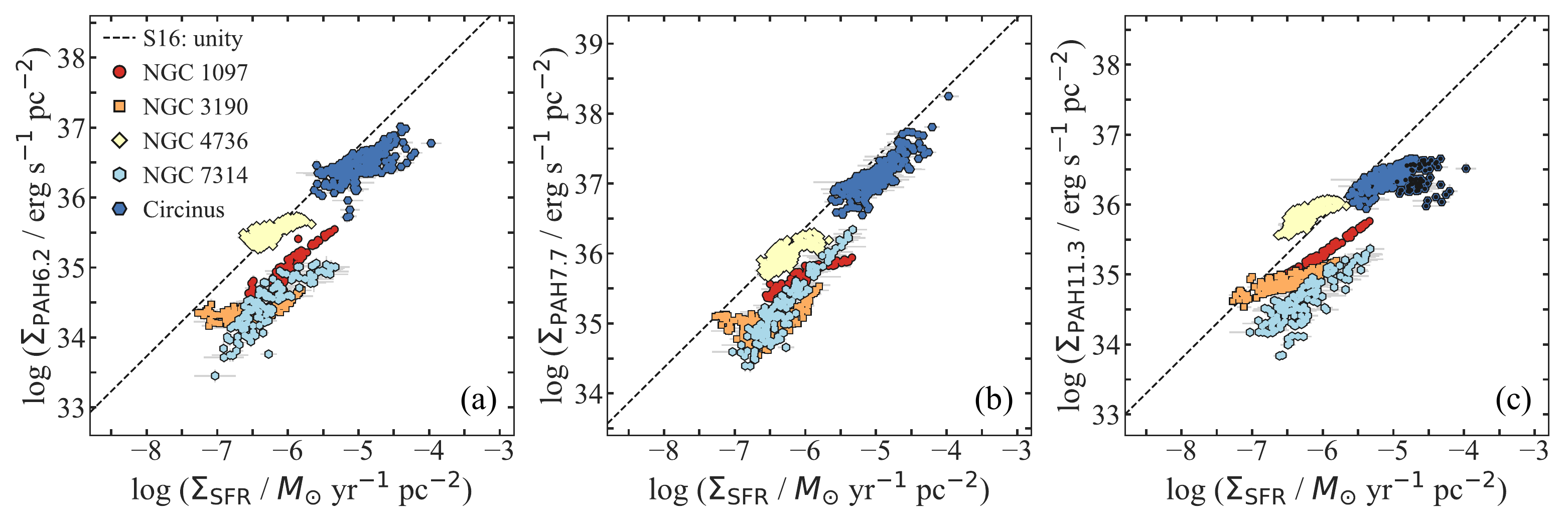}}
\caption{Correlations between the surface brightness of the PAH (a) 6.2\,$\mum$ (b) 7.7\,$\mum$, and (c) 11.3\,$\mum$ emission and the SFR surface density for $0\farcs2\times0\farcs2$ spaxels in the central $\sim 4\arcsec \times 4\arcsec$ regions of the five AGN targets, where $\Sigma_{\rm PAH} = L_{\rm PAH}\,cos(i)/A$ and $\Sigma_{\rm SFR} = {\rm SFR}\,cos(i)/A$. Here, $i$ is the galaxy inclination angle and $A$ is the spaxel area in $\rm pc^{2}$. The dashed lines represent the unity PAH-SFR relations calibrated by \citeauthor{Shipley.etal.2016} (\citeyear{Shipley.etal.2016}; S16: unity) for star-forming galaxies. The data points highlighted with black dots represent spaxels in the central $r \approx 15 - 25$ pc region of Circinus (the inner region is excluded from the analysis, as detailed in Section~\ref{sec2.2}). Unless otherwise specified, the analysis presented here and in subsequent figures includes only spaxels with measurements of all five H$_2$ transitions required for the shock-processing indicator and all three neon emission lines used for SFR estimation and the AGN-irradiation indicator.}\label{PAHvsSFR}
\end{figure*}

To compare PAH emission strengths among the five targets at fixed SFRs, we derived spaxel-based SFRs from the infrared neon lines [Ne~{\small II}]12.81$\mu$m, [Ne~{\small III}]15.56$\mu$m, and [Ne~{\small V}]14.32$\mu$m using the AGN-calibrated prescription of \cite{Zhuang&Ho2019}. This prescription explicitly accounts for the AGN contribution to the neon lines. As shown in Figure~\ref{PAHvsSFR}, spaxels in the central $\sim 4\arcsec \times 4\arcsec$ regions of the five AGN lie generally below the unity PAH-SFR relations calibrated by \cite{Shipley.etal.2016} for star-forming galaxies.\footnote{The unity relations of \cite{Shipley.etal.2016}, with the slope fixed at unity, were calibrated using Spitzer/IRS spectroscopy of 105 star-forming galaxies obtained over apertures larger than the pixel scales considered here. Nevertheless, these unity relations provide an appropriate baseline for the qualitative comparisons discussed here. They also derived linear calibrations for the PAH 6.2, 7.7, and 11.3 $\mum$ features, with slopes of 1.04, 1.00, and 0.94, respectively.} Despite the relative suppression of PAH emission across the sample, spaxels in NGC~4736 and Circinus display relatively enhanced PAH emission at a given SFR compared to those in the other three AGN. This trend does not hold in the inner regions of Circinus, particularly for PAH 11.3~$\mum$ feature. As shown in panel (c) of Figure~\ref{PAHvsSFR} (see also the resolved maps in Section~\ref{sec3.2}), this PAH feature exhibits pronounced suppression in the inner regions of Circinus, reaching levels comparable to those observed in NGC~7314. Such increasing suppression of PAH 11.3~$\mum$ feature toward the Circinus nucleus has also been reported by \citeauthor{Jensen.etal.2017} (\citeyear{Jensen.etal.2017}; see Figure~2 therein). The relatively stronger PAH emission in spaxels in NGC~4736 and Circinus outer-circumnuclear regions compared to the other three AGN is likely attributable to the (post-)starburst nature (\citealt{Pellegrini.etal.2002, Marconi.etal.1994}) of their central regions. In contrast, the relative suppression of PAH 11.3~$\mu$m feature in NGC~7314 and the inner regions of Circinus compared to the other three AGN reflects a reduced abundance of neutral PAHs, likely caused by the preferential ionization of PAHs in these environments, as discussed with model results in Section~\ref{sec3.2}.

As noted above, the relative suppression of intrinsic PAH emission in the vicinity of AGN is in theory attributed to the selective erosion and destruction of PAHs, primarily driven by AGN irradiation and shock processing that represent distinct modes of AGN feedback. Following this point, we performed a multiple linear regression analysis with the {\tt Python} package {\tt statsmodels} (\citealt{Seabold&Perktold2010}). This analysis aims to reproduce the observed PAH surface brightness (in unit of $\rm erg\,s^{-1}\,pc^{-2}$) as a function of SFR surface density (in unit of $\rm M_{\odot}\,pc^{-2}$), the ratio of the summed H$_2$\,$S(1)$ -- $S(5)$ emission to PAH 7.7\,$\mum$ feature (i.e., H$_2$\,$S(1–5)$/PAH7.7; hereafter $\rm rH_{2}$), and the ratio of [Ne~{\small V}]14.32$\mu$m emission to PAH 11.3\,$\mum$ feature (i.e., [Ne~{\small V}]14.32/PAH11.3; hereafter $\rm rNe$). The ratio of infrared H$_2$ transitions to PAH emission has long been proposed as a powerful tracer of shocks, as the intensities of infrared H$_2$ transitions generally scale closely with PAH emission over a wide range of radiation field strengths, whereas shock heating can produce an excess of H$_2$ emission (e.g., \citealt{Roussel.etal.2007, Ogle.etal.2010, Guillard.etal.2012, Riffel.etal.2020, Kristensen.etal.2023}). We adopt the ratio of the high-ionization [Ne~{\small V}] line to PAH emission as an indicator of AGN irradiation strength because the strength of this coronal line has been shown to correlate strongly with AGN luminosity over a wide range of luminosities and physical conditions. (e.g., \citealt{Weaver.etal.2010, Garcia-Bernete.etal.2017, Bierschenk.etal.2024, Annuar.etal.2025, Goold.etal.2026}). We note that the diagram consisting of these two ratios efficiently distinguish the two Seyfert galaxies  from the three LINERs studied here (see Figure~\ref{H2NeP} in Appendix~\ref{secA0}).

The multiple linear regression analysis yields
\noindent
\begin{align}\label{equpp}
\begin{aligned}
{\rm log}\,\Sigma_{\rm PAH}^{\rm fit}  = \alpha_{1}\,{\rm log}\,\Sigma_{\rm SFR} + \alpha_{2}\,{\rm log}\,{\rm rH_{2}} + \alpha_{3}\,{\rm log}\,{\rm rNe} + \beta,
\end{aligned}
\end{align}
\noindent
where the coefficients $\alpha_{1}$, $\alpha_{2}$, $\alpha_{3}$, and $\beta$ for PAH 6.2, 7.7, and 11.3\,$\mum$ features are listed in Table~\ref{tabcof1} in Appendix~\ref{secA0}. As shown in Figure~\ref{PAHSFR}, the multiple linear regression model incorporating SFR, $\rm rH_{2}$, and $\rm rNe$ reproduces the observed PAH emission well, with root-mean-square (RMS) scatters ($\sigma$ in Table~\ref{tabcof1}) of 0.25, 0.19, and 0.21 dex for PAH 6.2, 7.7, and 11.3\,$\mum$ features, respectively. These tight correlations support the interpretation that the general suppression of PAH emission in AGN, relative to that expected at a given SFR, is governed by the combined effects of AGN irradiation and shock processing.

\begin{figure*}[!ht]
\center{\includegraphics[width=1\linewidth]{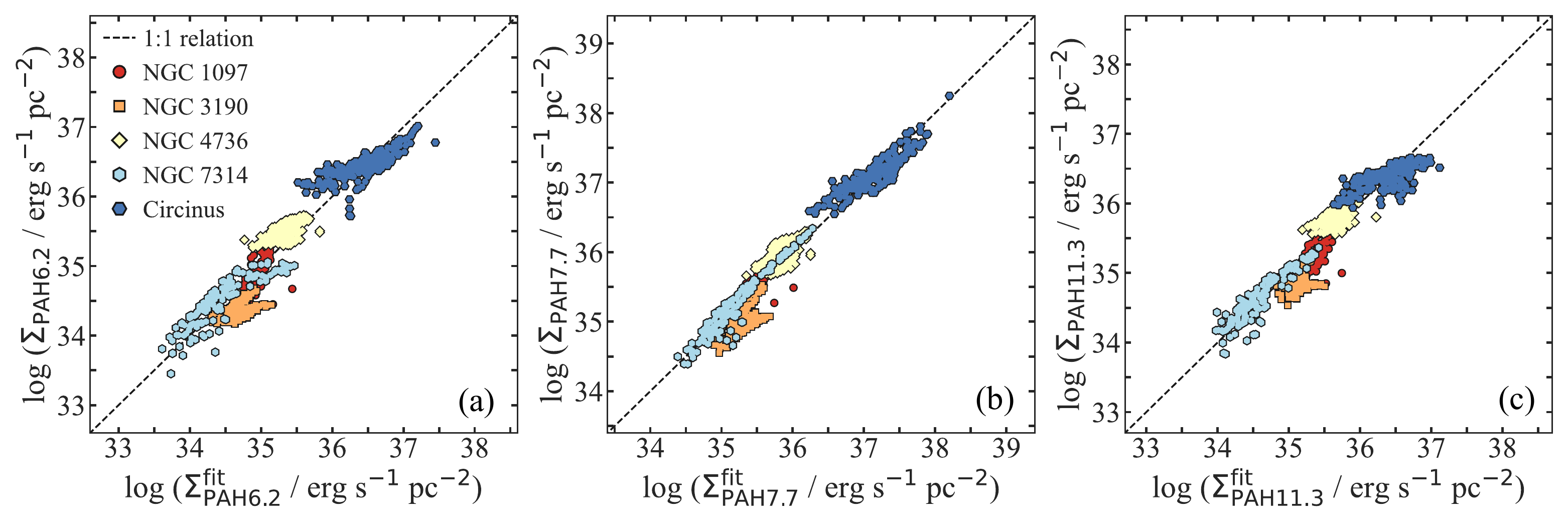}}
\caption{Correlations between the surface brightness of the measured PAH (a) 6.2\,$\mum$, (b) 7.7\,$\mum$, and (c) 11.3\,$\mum$ emission and those predicted by Equation~\ref{equpp} for data points in Figure~\ref{PAHvsSFR}, with the dashed line in each panel indicating the 1:1 relation.}\label{PAHSFR}
\end{figure*}

Intriguingly, the fitted correlation for PAH 7.7\,$\mu$m feature exhibits the smallest scatter among the three PAH features, suggesting that it provides the most robust tracer of SFR in the targets studied here. This result is noteworthy given the lack of consensus among previous studies on which PAH feature serves as the best SFR indicator, with some studies reporting consistent results and others reaching conflicting conclusions. Specifically, \cite{Diamond-Stanic&Rieke2010}, based on an analysis of 35 Seyfert nuclei on sub-kpc scales, argued that PAH 11.3\,$\mu$m feature, unlike PAH 7.7\,$\mu$m feature, remains a robust tracer of SFR--as indicated by the [Ne~{\small II}]12.81$\mu$m emission--even in nuclei exhibiting unusually high PAH 11.3/7.7 ratios. In contrast, \cite{LaMassa.etal.2012}, analyzing a larger sample of 264 star-forming galaxies, 51 composites, and 73 AGN over kpc scales, found that PAH 11.3\,$\mu$m feature is significantly suppressed relative to PAH 7.7\,$\mu$m feature in most of their AGN, raising concerns about its reliability as a standalone SFR indicator in such environments. Later, \cite{Esquej.etal.2014} analyzed the equivalent width of PAH 11.3\,$\mu$m feature (hereafter PAH EW11.3) in a sample of 29 Seyfert nuclei on sub-100-pc scales and found no significant dependence of PAH EW11.3 on AGN luminosity. They therefore concluded that there is no general evidence for suppression of PAH 11.3\,$\mu$m feature in the vicinity of AGN, supporting its use as an SFR proxy. Nevertheless, they also reported that PAH 11.3\,$\mu$m feature is undetected in the central $\sim$15 pc region of Circinus, corroborating the earlier findings of \cite{Roche.etal.2006} and consistent with the trend of increasing suppression of this feature in Circinus nucleus found here. Our results are broadly consistent with the conclusions of \cite{LaMassa.etal.2012}. Additional support comes from the linear PAH--SFR calibrations of \cite{Shipley.etal.2016}, which give a linear relation (slope = 1.0) for PAH 7.7\,$\mu$m feature but a slightly sublinear relation (slope = 0.94) for PAH 11.3\,$\mu$m feature. In Section~\ref{sec4}, we explore the physical mechanisms responsible for these disparate results and show that they can be explained by differences in sample selection within a unified framework that accounts for the combined effects of AGN irradiation and shock processing.

We additionally performed a complementary multiple linear regression analysis to calibrate the SFR surface density (in unit of $\rm M_{\odot}\,pc^{-2}$) as a function of PAH surface brightness (in unit of $\rm erg\,s^{-1}\,pc^{-2}$), $\rm rH_{2}$, and $\rm rNe$, thereby demonstrating the practical application of the above results for deriving SFRs from PAH emission in AGN. The complementary multiple linear regression analysis yields
\noindent
\begin{align}\label{equss}
\begin{aligned}
{\rm log}\,\Sigma_{\rm SFR} = \alpha_{1}\,{\rm log}\,\Sigma_{\rm PAH} + \alpha_{2}\,{\rm log}\,{\rm rH_{2}} + \alpha_{3}\,{\rm log}\,{\rm rNe} + \beta
\end{aligned}
\end{align}
\noindent
where the coefficients $\alpha_{1}$, $\alpha_{2}$, $\alpha_{3}$, and $\beta$ for PAH 6.2, 7.7, and 11.3\,$\mum$ features are listed in Table~\ref{tabcof2} in Appendix~\ref{secA0}. This multiple linear regression model yields PAH-based SFRs that are well correlated with the neon-based SFR, with RMS scatters of 0.24, 0.20, and 0.24 dex for PAH 6.2, 7.7, and 11.3\,$\mum$ features, respectively. By explicitly incorporating the effects of AGN irradiation and shock processing on PAH emission, this multiple linear regression approach offers a new method for calibrating SFR in AGN, warranting validation and refinement with larger samples in future studies.

\begin{figure*}[!ht]
\center{\includegraphics[width=0.95\linewidth]{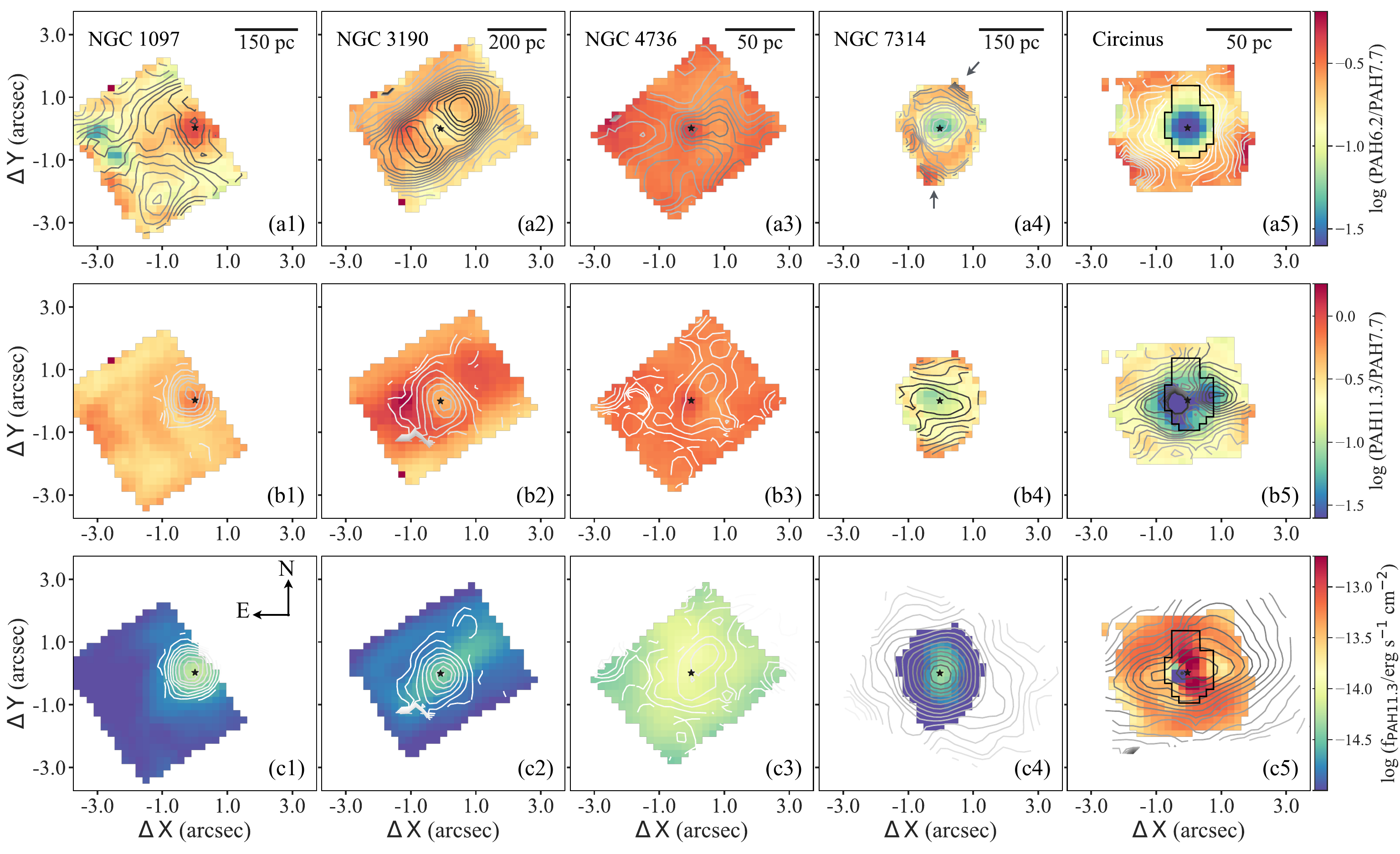}}
\caption{Spatially resolved maps of PAH 6.2/7.7 ratio (top panels), PAH 11.3/7.7 ratio (middle panels), and PAH 11.3 $\mum$ emission (bottom panels) for the five AGN targets. Each column corresponds to one target, with the AGN position denoted by a black star. A common color scale is adopted for all panels within a given row and is shown on the right. The contours in the top, middle, and bottom panels show the distributions of log\,($\rm rH_{2}$) ($-2.25$ to $-0.5$), log\,($\rm rNe$) ($-4.0$ to $0.15$), and log\,($f_{\rm [Ne~{\footnotesize V}]14.32{\rm \mum}}/{\rm erg\,s^{-1}\,cm^{-2}}$) ($-18.0$ to $-13.3$), respectively, with higher values indicated by darker colors on a common scale for each row. The black arrows in panel (a4) indicate the locations of the two radio spots in NGC~7314. The black polygons in Circinus outline the region within which the spaxels are shown here for illustrative purposes; these spaxels are excluded from the statistical analyses due to the lack of reliable spaxel-based measurements of [Ne~{\small III}]15.56$\mu$m and/or H$_2\,S(1)$ and $S(2)$, as detailed in Section~\ref{sec2.2}. North is up in all panels, and the physical scale of each target is labeled in the top-right corner of each top panel.}\label{PAHratioMaps}
\end{figure*}

\subsection{Correlation Between PAH Emission and AGN Activity}\label{sec3.2}

Beyond the intrinsic strength of individual PAH features, the observed ratios among different PAH features encode rich information about the size distribution and ionization state of PAH molecules, as well as the radiation field in which they are embedded (e.g., \citealt{Draine&Li2007, Draine.etal.2021, Rigopoulou.etal.2021, Rigopoulou.etal.2024}, and for reviews see \citealt{Tielens2008, Li2020}). In short, smaller PAHs emit more strongly at shorter (i.e., more energetic) wavelengths, as their lower heat capacity leads to higher excitation levels following the absorption of a UV photon. Neutral PAHs exhibit stronger 3.3 and 11.3 $\mum$ features due to more prominent C–H vibrational modes, whereas ionized (cationic) PAHs preferentially emit in the 6–9 $\mum$ range, where C–C vibrational modes dominate. Furthermore, a moderately stronger radiation field increases the PAH ionization fraction, and a harder, FUV-rich radiation field enhances the relative strength of the shorter-wavelength PAH features prior to their photo-destruction. In other words, variations in the observed PAH band ratios reflect changes in the underlying physical conditions, including those driven by different modes of AGN feedback.

\begin{figure*}[!ht]
\center{\includegraphics[width=0.95\linewidth]{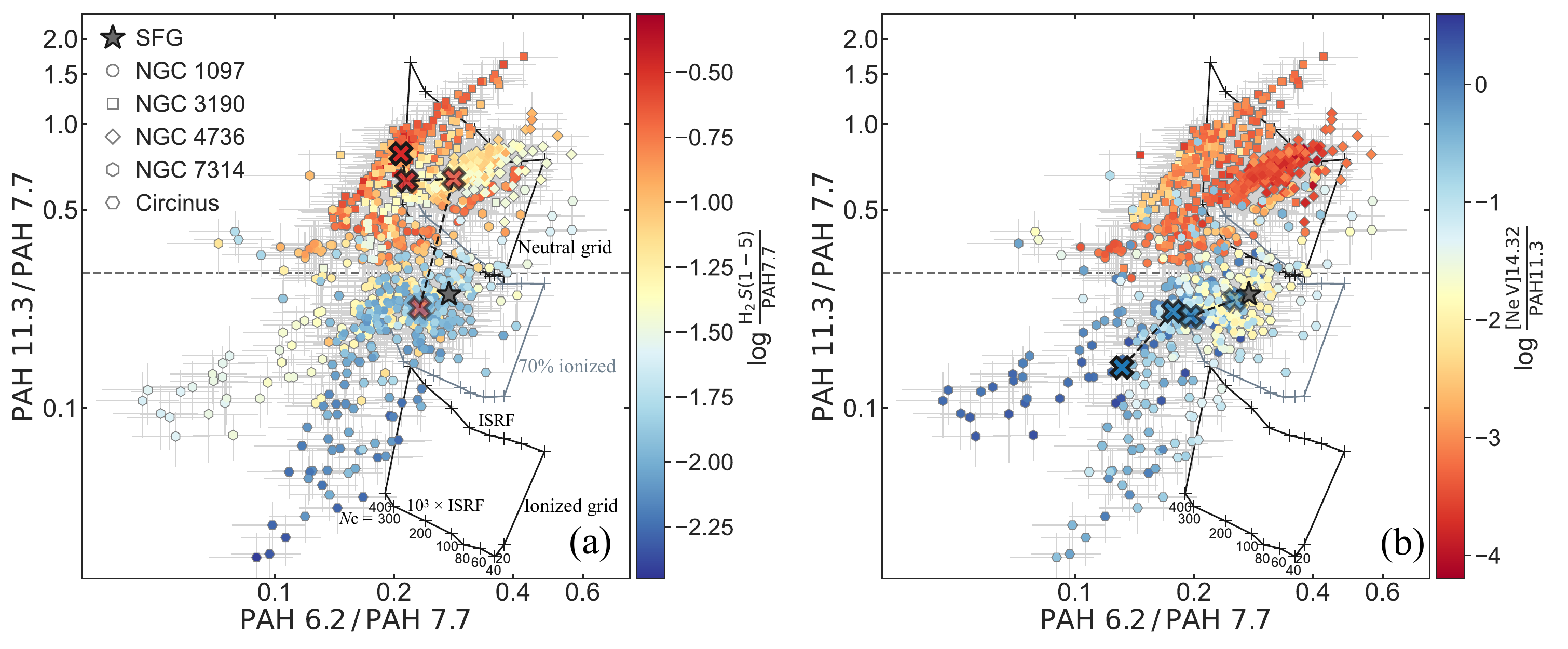}}
\caption{PAH band ratio diagrams for $0\farcs2\times0\farcs2$ spaxels in the central $\sim 4\arcsec \times 4\arcsec$ regions of the five AGN targets, color-coded by the strength indicators of (a) shock processing ($\rm rH_{2}$) and (b) AGN irradiation ($\rm rNe$), respectively. The horizontal dashed lines mark the empirical demarcation at PAH\,11.3/7.7 = 0.3. The reddish and bluish crosses in panels (a) and (b) show, for the five AGN targets, the median PAH band ratios of the spaxels in bins of increasing log\,($\rm rH_{2}$) of [$-2, -1.5$], [$-1.5, -1.0$], [$-1.0, -0.5$], and [$-0.5, 0$], and increasing log\,($\rm rNe$) of [$-2, -1.5$], [$-1.5, -0.75$], [$-0.75, 0$], and [$0, 0.5$] (from light to dark), respectively. The gray star labeled ``SFG'' represents the median PAH band ratios of the spatially resolved spaxels in star-forming galaxies measured by \cite{Zhang.etal.2022}; the individual SF spaxels, which are clustered around the median value, are omitted for clarity. The grids marked by `+' in each panel show model predictions of PAH band ratios from \cite{Rigopoulou.etal.2024} for neutral (top grids), 70\% ionized (middle grids), and fully ionized (bottom grids) PAHs of various sizes (carbon number $N_{\rm C} = 20-400$, increasing from the right to the left boundary of each grid). The grids span the interstellar radiation field (ISRF; top boundary) up to $10^{3}\times$ISRF (bottom boundary).}\label{PAHratios}
\end{figure*}

As presented in Figure~\ref{PAHratioMaps}, the spatially resolved maps of the two most widely used PAH 6.2/7.7 (top panels) and 11.3/7.7 (middle panels) ratios exhibit pronounced spatial variations, reflecting modifications in PAH properties toward the nucleus of each target. In particular, the nuclei of the two Seyfert galaxies studied here show systematically lower PAH 6.2/7.7 and 11.3/7.7 ratios ($<0.3$ and $<0.35$, respectively, at $r<1\arcsec$) than the three LINERs studied here. These lower ratios are plausibly driven by strong AGN irradiation, as indicated by the darker contours in panels (b4) and (b5) of Figure~\ref{PAHratioMaps}, which trace regions of elevated $\rm rNe$ values. In contrast, the comparatively higher PAH 11.3/7.7 ratios observed in the three LINERs are more likely associated with shock processing, as suggested by the darker contours in panels (a1)$-$(a3) of Figure~\ref{PAHratioMaps}, which trace regions of elevated $\rm rH_{2}$ values (see also \citealt{Zhang.etal.2026} for a detailed discussion of the role of shock processing in these LINERs). As discussed below and further elaborated in Section~\ref{sec4}, this difference is not merely a consequence of AGN type, but instead reflects the distinct physical conditions driven by different modes of AGN feedback. We also note that, although NGC~3190 exhibits overall large PAH 6.2/7.7 and 11.3/7.7 ratios, its nucleus shows a modest decrease in these ratios. This behavior is likely driven by the combined effects of the spatially concentrated AGN-irradiation-dominated region and relatively weak shock processing around the nucleus, as evidenced by the nuclear $\rm rNe$ peak and $\rm rH_{2}$ valley in panels (b2) and (a2) of Figure~\ref{PAHratioMaps}.

Other noteworthy features in Figure~\ref{PAHratioMaps} include compact structures in NGC~1097 that exhibit lower PAH 6.2/7.7 ratios toward the east (see the greenish regions in panel a1). These structures are also evident in low-ionization emission line maps (e.g., [Ar~{\small II}]6.98$\mu$m, [Ne~{\small II}]12.81$\mu$m), but are absent in higher-ionization (e.g.,  [Ne~{\small III}]15.56$\mu$m, [O~{\small IV}]25.89$\mu$m) and $\rm H_{2}$ emission line maps not shown here. In NGC~7314, the two regions at the northern and southern tips that show modestly enhanced shock processing, i.e., higher $\rm rH_{2}$ values as shown in panel (a4), are spatially associated with two radio spots detected by \cite{Thean.etal.2000}. Additionally, the tilted east-west elongated regions exhibiting strong AGN irradiation, i.e., higher $\rm rNe$ values as shown in panel (b4), align with the orientation of its AGN ionization cone, which has a wide opening angle as revealed by high-ionization emission-line maps (see e.g., panel c4 and also \citealt{daSilva.etal.2023}). In contrast, the modest north-south elongation of the circumnuclear PAH emission in NGC~7314 (see panel c4) is spatially coincident with the circumnuclear cold gas distribution and aligned with the inner dusty molecular torus (\citealt{Garcia-Burillo.etal.2021}). In Circinus, the regions with low PAH 11.3/7.7 ratios also coincide with the ionization cone traced by high-ionization emission-line maps (see e.g., panel c5 and also \citealt{Marconi.etal.1994}), as well as the polar dust elongation (\citealt{Izumi.etal.2018, Lopez-Rodriguez.etal.2026}); however, note that the PAH measurements may be biased by the strong continuum of this target in the circumnuclear region outlined by the black polygons in the rightmost panels. These spaxels were therefore excluded from all statistical analyses presented in this study.

\begin{figure*}[!ht]
\center{\includegraphics[width=0.95\linewidth]{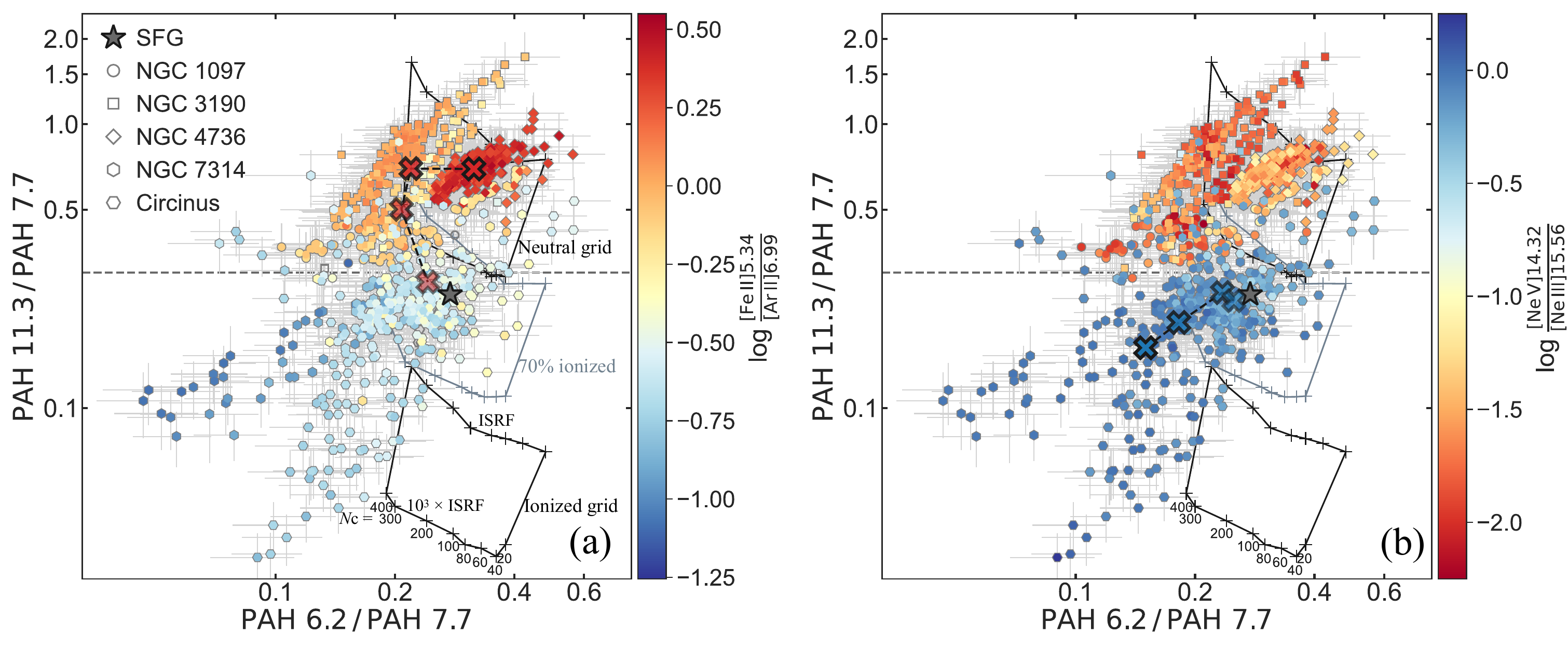}}
\caption{Same as Figure~\ref{PAHratios} but color-coded by the different strength indicators of (a) shock processing ([Fe~{\footnotesize II}]5.34/[Ar~{\footnotesize II}]6.99) and (b) AGN irradiation ([Ne~{\footnotesize V}]14.32/[Ne~{\footnotesize III}]15.56). The reddish and bluish crosses in panels (a) and (b) show, for the five AGN targets, the median PAH band ratios of the spaxels in bins of increasing log\,([Fe~{\footnotesize II}]5.34/[Ar~{\footnotesize II}]6.99) of [$-0.5, -0.25$], [$-0.25, 0$], [$0, 0.25$], and [$0.25, 0.5$], and increasing log\,([Ne~{\footnotesize V}]14.32/[Ne~{\footnotesize III}]15.56) of [$-0.5, -0.3$], [$-0.3, -0.15$], [$-0.15, 0$], and [$0, 0.2$] (from light to dark), respectively.}\label{PAHratiosII}
\end{figure*}

In the following, we focus on the physical insights provided by the PAH band ratios of the spaxels in the five AGN, in combination with the diagnostic ratios $\rm rH_{2}$ and $\rm rNe$. As shown in Figure~\ref{PAHratios}, the spaxels in the five AGN overall exhibit lower PAH 6.2/7.7 ratios compared to the spaxels in star-forming galaxies. More importantly, the spaxels in the targets studied here exhibit a bimodal distribution of PAH 11.3/7.7 ratios (with PAH 11.3/7.7 = 0.3 serving as a useful demarcation), as already seen in their spatially resolved maps. Relative to spaxels in star-forming galaxies, the spaxels in the LINERs and Seyfert galaxies studied here display systematically higher and lower PAH 11.3/7.7 ratios, respectively. As indicated by the color coding, this bimodal distribution appears to arise from the combined influence of enhanced shock processing in the three LINERs and stronger AGN irradiation in the two Seyfert galaxies, with star formation activity providing the underlying baseline. As outlined in Section~\ref{sec3.1} and discussed further below, the lower PAH 11.3/7.7 ratios in the two Seyferts are primarily driven by the relative suppression of their PAH 11.3 $\mum$ emission, reflecting a lower fraction of neutral PAHs. To highlight these trends, we plot for the five AGN the median PAH band ratios of the spaxels in bins of increasing log\,($\rm rH_{2}$) and log\,($\rm rNe$) as reddish and bluish crosses (from light to dark) in panels (a) and (b) of Figures~\ref{PAHratios}, respectively. These reddish and bluish crosses clearly trace two distinct trends of progressively stronger PAH modifications associated with enhanced shock processing and stronger AGN irradiation, respectively. Note again that this difference is not merely a consequence of AGN type, but instead reflects the distinct physical conditions associated with the two modes of AGN feedback.

The model grids in Figure~\ref{PAHratios} confirm that the systematically higher and lower PAH 11.3/7.7 ratios observed in the three LINERs and the two Seyfert galaxies, respectively, correspond to larger fractions of neutral and ionized PAHs, while the generally lower PAH 6.2/7.7 ratios indicate relatively larger PAH sizes. As discussed by \cite{Zhang.etal.2026}, a natural explanation for this bimodal distribution is that shock processing, if present, preferentially destroy both smaller and more vulnerable ionized PAHs (\citealt{Allain.etal.1996, Holm.etal.2011}), leading to a relative enhancement of larger, neutral PAHs. In contrast, intense AGN irradiation, as observed in the two Seyfert galaxies studied here, preferentially ionizes PAHs and subsequently destroys most of the smaller grains, resulting in a relative enhancement of larger, ionized PAHs. In particular, for spaxels in NGC~7314, which exhibits both strong AGN irradiation and moderate shock processing, the bimodal distribution in their PAH band ratios is most consistent with this scenario, as further elaborated in Section~\ref{sec4}. Direct destruction of PAHs by hard X-ray photons, without prior ionization, may mimic the effects of shock processing (\citealt{Micelotta.etal.2010b}), highlighting the need for dedicated studies of X-ray-rich environments around more luminous AGN. Partially destroyed PAHs with open or irregular structures and increased hydrogenation (e.g., catacondensed PAHs) may exhibit enhanced PAH 11.3/7.7 ratios (e.g., \citealt{Pantoni.etal.2026, Li2020}), although this interpretation remains debated (\citealt{Rigopoulou.etal.2024}).

To further confirm the two distinct trends of progressive PAH modifications, we turn to additional independent diagnostics based on infrared ionized emission lines. Here, we use the [Ne~{\small V}]14.32/[Ne~{\small III}]15.56 ratio as an additional proxy for AGN irradiation, as it is less susceptible to star formation activity and correlates well with the ionization parameter $U$ in AGN (\citealt{Zhang.etal.2025}). In addition, we adopt the [Fe~{\small II}]5.34/[Ar~{\small II}]6.99 ratio as a proxy for shock processing, which is sensitive to shocks because fast shocks can efficiently liberate iron from dust grains while having little effect on the noble gas argon. Previous studies have shown that enhanced [Fe~{\small II}] emission in AGN is closely associated with shocked regions (e.g., \citealt{Forbes&Ward1993, Mouri.etal.2000, Storchi-Bergmann.etal.2009, Colina.etal.2015, Koo.etal.2016, Rodriguez-Ardila.etal.2017, Durre&Mould2018, Riffel.etal.2021, Costa-Souza.etal.2026}). The diagrams consisting of these ionized emission-line ratios efficiently distinguish the two Seyfert galaxies  from the three LINERs (see Figure~\ref{LineD} in Appendix~\ref{secA0}): the former are dominated by AGN excitation, whereas the latter are better explained by shock excitation. Correspondingly, we revisit Figure~\ref{PAHratios}, but now color-code the data points using these ionized emission-line ratio diagnostics. As shown in Figure~\ref{PAHratiosII}, the two distinct trends of progressive PAH modifications remain evident, although some discrepancies between Figures~\ref{PAHratios} and \ref{PAHratiosII} are also present, particularly in panel (a).

In contrast to Figure~\ref{PAHratios}(a), Figure~\ref{PAHratiosII}(a) exhibits a different final direction of the red crosses, driven by the highly elevated [Fe~{\small II}]5.34/[Ar~{\small II}]6.99 ratios in NGC~4736. A similar elevation is plausibly also present in Circinus, with NGC~7314 serving as a reference. This result is not surprising, as the shocks that enhance H$_2$ emission are not necessarily the same as those responsible for enhancing [Fe~{\small II}] emission (e.g., \citealt{Hollenbach&McKee1989, Mouri.etal.2000, Allen.etal.2008, Kristensen.etal.2023}); in particular, the fast shocks (with velocities of hundreds of $\rm km\,s^{-1}$) required to produce ionized gas emission would instead lead to H$_2$ dissociation. As discussed by \cite{Riffel.etal.2026a}, H$_2$ emission is more likely to originate in post-shock gas, where molecules reform after being dissociated by the fast shocks responsible for ionized gas emission (e.g., \citealt{Guillard.etal.2009, Richings&Faucher-Giguere2018a, Richings&Faucher-Giguere2018b}). Again, the difference between Figures~\ref{PAHratios}(a) and \ref{PAHratiosII}(a) is likely related to the (post-) starburst nature of the central regions in NGC~4736 and Circinus, given that both X-ray irradiation and shock heating can contribute to [Fe~{\small II}] emission (e.g., \citealt{Mouri.etal.2000, Allen.etal.2008, Dors.etal.2012}). Specifically, a cluster of at least four bright point-like X-ray sources is present in the central $r \leq 5\arcsec$ region of NGC~4736, with the brightest ones attributed to gas heated by supernovae rather than the low-luminosity AGN at its center (see \citealt{Pellegrini.etal.2002}). Despite the differences, Figures~\ref{PAHratios} and \ref{PAHratiosII} consistently support the existence of two distinct AGN feedback modes: one associated with shock processing and the other with AGN irradiation, although supernovae in (post-)starburst environments may also contribute to shock processing and the resulting PAH modifications.

\begin{figure}[!t]
\center{\includegraphics[width=0.95\linewidth]{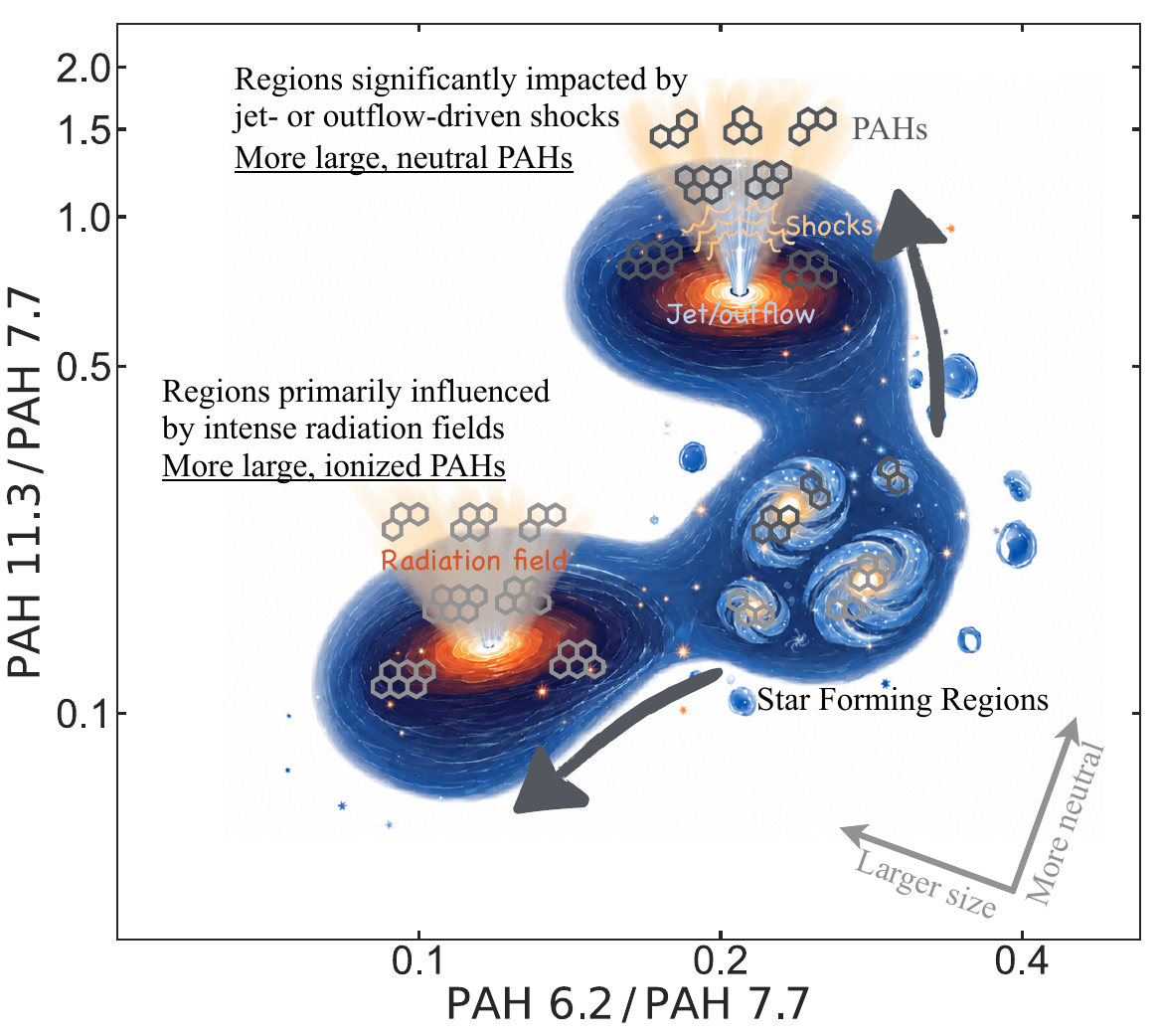}}
\caption{Illustrative schematic showing the two regimes in our AGN targets affected by the two feedback modes associated with AGN irradiation and shock processing, respectively. Gray arrows in the lower right corner indicate the directions in theory of increasing overall PAH size and neutral fraction as the PAH band ratios change. Using star-forming regions as a baseline, AGN irradiation tends to drive the heating and ionization of PAH molecules, followed by the preferential photodestruction of smaller PAHs, leading to a population dominated by larger and more ionized PAHs. In contrast, shock processing appears to preferentially destroy ionized PAHs, resulting in a higher survival fraction of large and neutral PAHs.}\label{Illustration}
\end{figure}

In summary, regardless of the diagnostics employed, the spaxels in the five AGN consistently reveal two clearly distinguishable trends of progressive PAH modification associated with the two AGN feedback modes. We also confirm that the bimodal distribution and the two distinct trends discussed above persist when the y-axis is replaced by the PAH 11.3/6.2 ratio (see Figure~\ref{PAHratios_apx} in Appendix~\ref{secA0}), which also traces PAH ionization and is less sensitive to continuum contamination. In addition, the PAH band ratios derived from integrated spectra extracted within large apertures, $r \leq 0\farcs5$, $0\farcs5 \leq r \leq 1\farcs0$, and $1\farcs0 \leq r \leq 1\farcs5$, exhibit the the same bimodal distribution as those observed among the spaxels in the five AGN (see the black-outlined points in Figure~\ref{PAHratios_apx}), demonstrating that these trends are robust against potential PSF-related artifacts. This result is expected, as all spectral data cubes were convolved to a common PSF prior to the spaxel-based measurements. Also note that all targets studied here exhibit prominent PAH features in the JWST/MRS spectra (see Figure~\ref{Fit_Demos} and Table~\ref{tabflux} in Appendix~\ref{secA0} for examples); therefore, the two distinct trends are unlikely to arise from noise-induced artifacts. In short, the two distinct trends observed among the spaxels across the five targets naturally reflect the two primary modes of AGN feedback, as illustrated in Figure~\ref{Illustration} and discussed in detail in Section~\ref{sec4}.

Additionally, we note that different analysis methods (e.g., {\tt PAHFIT}, \citealt{Smith.etal.2007}; {\tt CAFE}, \citealt{Marshall.etal.2007}; {\tt SPIRIT}, \citealt{Donnan.etal.2024}; {\tt PAHdb}, \citealt{Maragkoudakis.etal.2025}) can yield systematically different PAH measurements from JWST spectra. The resulting differences in derived PAH fluxes between {\tt CAFE} and {\tt SPIRIT} can commonly reach a factor of $\sim$2, with {\tt SPIRIT} generally yielding lower PAH fluxes (\citealt{Pantoni.etal.2026}). In addition, decompositions based on {\tt PAHFIT} and {\tt CAFE} differ from {\tt PAHdb} modeling by $\sim$7\% in the inferred PAH ionization fraction (\citealt{Maragkoudakis.etal.2025}). Specifically, for spectra extracted from the $r = 0\farcs5 - 1\farcs0$ apertures in the five targets, {\tt SPIRIT} yields PAH 6.2/7.7 and 11.3/7.7  ratios that are systematically more than $\sim$20\% higher, respectively, than those derived using the method adopted in this work (which is essentially equivalent to {\tt PAHFIT}), with differences exceeding 100\% in some cases. The differences in the measurements primarily arise from the different treatments of extinction and dust continuum, which consequently lead to systematic differences in the derived PAH measurements (see also discussion in \citealt{VanDePutte.etal.2025}). We therefore caution readers that systematic differences among analysis methods may lead to different interpretations of the observations. Nevertheless, the two distinct trends of progressive PAH modification observed across the five targets presented here are consistently supported by multiple diagnostics presented above, together with the additional evidence discussed in Section~\ref{sec4}. This consistency strongly suggests that these trends are robust and reflect genuine physical differences rather than methodological artifacts.

\section{Implications for AGN Feedback Modes}\label{sec4}

\subsection{Coexistence of Kinetic and Radiative AGN Feedback}\label{sec4.1}

As reported by \cite{Zhang.etal.2026}, the H$_2$ transitions in the nuclear regions of the three LINERs studied here are not fully thermalized, with slow, plausibly jet-driven molecular shocks associated with AGN providing the additional excitation. Specifically, matching the observed H$_2$ line ratios, $S(3)$/$S(1)$ and 1-0\,$O(5)$/$S(3)$, to the predictions of the theoretical models of \cite{Kristensen.etal.2023} indicates the presence of C-type molecular shocks with velocities of $v_{\rm s} \lesssim \rm 10\,km\,s^{-1}$. Moreover, they found that coupling only $\sim0.1-1$\% of the AGN jet power to the surrounding gas is sufficient to reproduce the observed H$_2$ fluxes in the nuclear regions of the three LINERs. Consistently, growing evidence suggests that the central engines of low-luminosity AGN are often dominated by jet power (e.g., \citealt{Fernandez-Ontiveros.etal.2012, Goold.etal.2026}), producing radio jets that span pc to kpc scales, depending on the jet power (\citealt{Mezcua&Prieto2014}). In contrast, Figure~\ref{LineD} shows that the radiative shocks traced by ionized emission lines have much higher velocities, with $v_{\rm s} \approx 100-400\,\rm km\,s^{-1}$, consistent with the shock velocity range of $\sim200-300\,\rm km\,s^{-1}$ modeled by \cite{Dopita.etal.2015} for the prototypical LINER NGC~1052. As discussed by \cite{Goold.etal.2024}, the fast radiative shocks observed in the low-luminosity AGN are likely driven by outflows powered by radiatively inefficient accretion flows (RIAFs) or by radio jets. The distinct shock velocities traced by the H$_2$ transitions and ionized emission lines in the three LINERs are consistent with the broader line widths and blueshifted ionized emission-line profiles observed in their nuclear regions (\citealt{Zhang.etal.2026}).

PAH emission primarily arises from photodissociation regions adjacent to warm molecular gas (e.g., \citealt{Chown.etal.2025, Maragkoudakis.etal.2026}) and is well correlated with the cold molecular gas component (e.g., \citealt{Cortzen.etal.2019, Leroy.etal.2023, Zhang&Ho2023a}). In addition, even large PAHs with $N_{\rm C}=200$ are expected to be destroyed by shocks with $v_{\rm s} \geq 125\,\rm km\,s^{-1}$ (\citealt{Micelotta.etal.2010a}). Given the tight correlation between PAH emission and the cold gas component, as well as the survival of PAHs with $N_{\rm C}\leq200$ in regions dominated by shock processing across the five targets (Figures~\ref{PAHratios} and \ref{PAHratiosII}), the observed shock processing of PAHs is likely dominated by low-velocity molecular shocks, while high-velocity radiative shocks may also play a significant role. Conservatively, the shocks responsible for modifying PAH properties are not necessarily limited to velocities of $\lesssim10\,\rm km\,s^{-1}$ but are likely to have velocities below $100\,\rm km\,s^{-1}$. Additional evidence for the importance of low-velocity shocks comes from the temperature distributions of warm molecular gas. Under local thermodynamic equilibrium, the relative populations of H$_2$ pure rotational transitions can be described by a power-law temperature distribution, $dN=mT^{-\beta}dT$, where lower values of $\beta$ indicate a larger fraction of warmer molecular gas and thus stronger heating by the underlying feedback processes. Following the methodology of \cite{Togi&Smith2016}, we derived $\beta$ values across the fields of view of the five targets by fitting the six H$_2$ pure rotational transitions, $S(1)-S(6)$. We find that shock-dominated regions exhibit $\beta \gtrsim 4.6$, whereas irradiation-dominated regions have $\beta \lesssim 4.6$, with the latter consistent with the range commonly observed in Seyfert nuclei (e.g., \citealt{Zhang&Ho2023c, Davies.etal.2024, Delaney.etal.2026}). The larger $\beta$ values in shock-dominated regions are consistent with low-velocity shocks that primarily heat the lower-excitation H$_2$ rotational levels. In contrast, sufficiently strong shocks, same as AGN irradiation, can populate higher-$J$ H$_2$ rotational levels, producing flatter temperature distributions and consequently lower $\beta$ values. 

As shown in Figure~\ref{PAHratioMaps}, the spaxels in the innermost region of NGC~7314 exhibit PAH band ratios indicative of a relatively high abundance of large PAHs and a high, though not extreme, ionization fraction. These results likely reflect the combined effects of the intense AGN irradiation in this target, which tends to ionize all and destroy smaller PAHs, together with modest shock processing, which preferentially enhances the destruction of fully ionized PAHs. The presence of notably large and ionized PAHs in the nuclear region of NGC~7314 is also reported by \citeauthor{Diamond-Stanic&Rieke2010} (\citeyear{Diamond-Stanic&Rieke2010}; see Figure~8 therein). Their measurements of PAH~6.2/7.7 $\approx 0.14$ and 11.3/7.7 $\approx 0.21$ from Spitzer/IRS spectroscopy place NGC~7314 at the extreme left of their PAH~11.3/7.7 vs PAH~6.2/7.7 diagram, distinguishing it from the nuclei of other Seyfert galaxies in their sample. More intriguingly, in the nuclear region of NGC~7314 examined here, larger and more highly ionized PAHs preferentially trace the AGN ionization cone (Section~\ref{sec3.2}), whereas larger but weakly ionized, and even neutral, PAHs are predominantly found along its periphery, particularly in regions spatially associated with the two radio spots that exhibit evidence of enhanced shock processing. Specifically, spaxels in NGC~7314 with PAH 11.3/7.7 $>0.4$ (predominantly neutral PAHs) exhibit log\,($\rm rH_{2}$) and log\,($\rm rNe$) values of $\sim-1.15$ to $-0.65$ and $\sim-1.05$ to $-0.40$, respectively (see Figure~\ref{H2NeP}). By contrast, spaxels in NGC~7314 with PAH 11.3/7.7 $<0.2$ and 6.2/7.7 $<0.15$ (predominantly ionized PAHs) show values that fall below and above these ranges, respectively, namely $\sim-1.60$ to $-1.30$ (relatively weaker shock processing) and $\sim-0.10$ to $0.45$ (relatively stronger AGN irradiation). This trend is qualitatively the same for the diagnostics based solely on ionized-emission lines (see Figure~\ref{LineD}). Notably, all the spaxels in NGC~7314 with PAH 11.3/7.7 $>0.4$ have log\,($\rm rH_{2}$) values above the threshold for a significant shock contribution (i.e., $-1.45$; \citealt{Guillard.etal.2012, Zhang.etal.2026}), highlighting the importance of shock processing in shaping PAH properties even in Seyfert galaxies.

Furthermore, according to \cite{Garcia-Bernete.etal.2024b}, JWST spectroscopy shows that the arcsecond-scale nuclear regions of three Seyfert galaxies with log\,$L_{\rm bol}/{\rm erg\,s^{-1}} \approx 44.2 - 44.3$, as well as their off-nuclear regions along the projected directions of AGN-driven outflows, exhibit elevated PAH 11.3/7.7 ratios and reduced PAH 6.2/7.7 ratios. Their PAH measurements (PAH 11.3/7.7 $\sim 0.5$ and PAH 6.2/7.7  $\sim 0.1-0.4$) in these regions are consistent with the values of the NGC~7314 spaxels located at the periphery of the AGN ionization cone, particularly those in regions exhibiting stronger shock processing. Importantly, \cite{Garcia-Bernete.etal.2024b} also reported elevated H$_2$/PAH ratios in the nuclear regions of these Seyfert galaxies and along their projected outflow directions, which coincide with their radio jet axes (\citealt{Zhang.etal.2024a}). More quantitatively, \cite{Zhang.etal.2024b} showed with JWST spectroscopy of another three Seyfert galaxies (log\,$L_{\rm bol}/{\rm erg\,s^{-1}} \approx 43.4 - 44.3$) that the $3\arcsec\times3\arcsec$ apertures with PAH 11.3/7.7 $>0.4$ exhibit log\,($\rm rH_{2}$) values of $\sim-1.00$ to $-0.05$, comparable to those observed in the three LINERs discussed here (see Figure~6 in \citealt{Zhang.etal.2026}). In contrast, the central $3\arcsec\times3\arcsec$ aperture of MCG-05-23-016 (PAH 11.3/7.7 $=0.37$), the only one among the three Seyfert galaxies where radiative effects play an important role as discussed therein, exhibits a comparatively lower log\,($\rm rH_{2}$) value of $-1.27 \pm 0.18$. In addition, according to \cite{Zhang&Ho2023c}, JWST spectroscopy shows that most 100-pc-scale spaxels in the circumnuclear region of NGC~7469 (log\,$L_{\rm bol}/{\rm erg\,s^{-1}} \approx 44.5$) exhibit PAH band ratios consistent with star-forming regions, particularly those located on the nuclear starburst ring; however, the innermost spaxels in NGC~7469 show elevated PAH 11.3/7.7 ratios of $\sim 0.4 - 0.5$, along with signatures of outflows. In particular, spaxels in the innermost 200 pc of NGC~7469 exhibit log\,($\rm rH_{2}$) and log\,($\rm rNe$) values of $-1.35$ and $-0.80$, respectively, consistent with a substantial shock contribution and comparatively weak AGN irradiation. These results further suggest that shock processing plays an important role in shaping the observed PAH properties even in Seyfert galaxies. However, we caution that direct destruction of PAHs by X-ray photons may also contribute in these higher-luminosity Seyfert nuclei, as noted in Section~\ref{sec3.2}.

\subsection{A Practical Unified Framework for AGN Feedback}\label{sec4.2}

Considered together, the JWST measurements in the literature and the spatially resolved results presented here suggest that, when AGN irradiation is the dominant feedback mechanism, it primarily modifies PAHs through radiative heating and ionization while preferentially photo-destroying the smaller PAHs. In the five targets examined here, this mode of feedback appears to be largely associated with AGN ionization cones. This scenario is consistent with the findings of \cite{Xie&Ho2022}, who reported elevated fractions of large and ionized PAHs on global scale (i.e., $\sim$ 10 kpc) in a sample of 86 low-redshift ($z \approx 0.05-0.4$) quasars with log\,$L_{\rm bol}/{\rm erg\,s^{-1}} \geq 44.5$, a regime in which radiative-mode feedback is expected to dominate globally. In contrast, when shock processing--driven by either jets or outflows in diverse AGN environments--becomes the dominant or a significant component of the feedback process, ionized PAHs are preferentially destroyed through shock interactions, thereby increasing the relative abundance of the surviving neutral PAHs. This mode of feedback appears to operate over more extended spatial scales in the targets examined here and is observed across a wider range of AGN types and luminosities in the literature. In particular, shocks arise from jets are more commonly associated with LINERs (see e.g., \citealt{Zhang.etal.2026}), whereas shocks driven by outflows are also frequently observed in Seyfert galaxies (see e.g., \citealt{Garcia-Bernete.etal.2024b}). These results are consistent with the conclusions of \cite{Zhang.etal.2022}, who showed that the relatively low PAH 6.2/7.7 and elevated PAH 11.3/7.7 ratios observed in spatially resolved 10\arcsec\ scale spaxels across a sample of 30 nearby AGN with log\,$L_{\rm bol}/{\rm erg\,s^{-1}} \approx 39.0-42.5$ cannot be attributed solely to the AGN radiation field, but instead point to the combined influence of AGN radiation fields and shocks. The above results point to a unified framework in which AGN type is not the primary factor governing the impact of AGN feedback. Rather, the underlying physical conditions--specifically the relative importance of AGN irradiation and shock processing--play a more fundamental role in determining the effectiveness of AGN feedback, and in shaping its influence on the ISM. Consistent with the unified framework proposed here, a growing number of studies combining (rest-frame) optical IFU spectroscopy with radio observations have revealed compelling evidence for the coexistence of radiative and kinetic feedback in AGN environments (e.g., \citealt{Rupke&Veilleux2011, Girdhar.etal.2022, Venturi.etal.2023, Roy.etal.2026}). Collectively, these findings support a unified view of AGN feedback as a multi-channel process, in which multiple feedback mechanisms can operate simultaneously, rather than being treated as mutually exclusive modes.

Within this unified framework, the seemingly disparate results discussed in Section~\ref{sec3.1} can be reconciled by accounting for the differences in sample selection. Specifically, the local ($z \lesssim 0.015$) Seyfert nuclei analyzed by \cite{Diamond-Stanic&Rieke2010} have PAH 11.3/7.7 $\approx 0.2-1.0$ and exhibit a significant positive correlation between the PAH 11.3/7.7 ratio and the H$_2$\,$S(3)$/(PAH7.7+PAH11.3) ratio (see Figure 6 therein). Remarkably, the six nuclei with the highest H$_2$\,$S(3)$/(PAH7.7+PAH11.3) ratios and PAH 11.3/7.7 $\gtrsim 0.6$ deviate from the PAH7.7--[Ne~{\small II}]12.81 correlation, yet still follow the corresponding PAH11.3--[Ne~{\small II}]12.81 correlation. This behavior led \cite{Diamond-Stanic&Rieke2010} to conclude that the 11.3\,$\mu$m PAH feature provides a more robust SFR indicator in these systems. In contrast, the slightly distant ($z \approx 0.025- 0.1$) AGN sample presented by \cite{LaMassa.etal.2012} exhibits generally lower PAH 11.3/7.7 ratios ($\sim 0.1 - 0.4$) and a clear negative correlation between the PAH EW11.3, as well as the PAH 11.3/7.7 ratio, and the [Ne~{\small V}]14.32/[Ne~{\small II}]12.81 ratio (see Figures 7 and 10 therein). Furthermore, the PAH EW11.3 values ($\lesssim 0.35$) of the local Seyfert nuclei studied by \cite{Esquej.etal.2014} lie within the regime where the negative [Ne~{\small V}]14.32/[Ne~{\small II}]12.81--PAH EW11.3 correlation reported by \cite{LaMassa.etal.2012} starts to flatten at the high [Ne~{\small V}]14.32/[Ne~{\small II}]12.81 end. Within the unified framework, the AGN samples studied by \cite{Diamond-Stanic&Rieke2010} and \cite{LaMassa.etal.2012} predominantly trace the two feedback regimes dominated by shock processing and AGN irradiation, respectively, with the former preferentially producing weaker PAH 7.7\,$\mu$m feature and the latter preferentially producing weaker PAH 11.3\,$\mu$m feature. Moreover, the generally higher redshifts and larger fraction of ionized PAHs in the AGN sample presented by \cite{LaMassa.etal.2012} are consistent with stronger radiative-mode AGN feedback at higher redshifts. Supporting this interpretation, \cite{Lofaro.etal.2026} recently found from their JWST spectroscopy that the only AGN-hosting galaxy in the PAHSPECS sample of five star-forming galaxies at $z \approx 1.1$ exhibits PAH 6.2/7.7 $\approx 0.13$ and 11.3/7.7 $\approx 0.16$ (extinction corrected), indicative of radiative-mode AGN feedback. In contrast, the four purely star-forming galaxies have PAH 6.2/7.7 $\approx 0.33$ and 11.3/7.7 $\approx 0.1$--$0.3$, values more consistent with those measured in the resolved star-forming spaxels of local galaxies (see also \citealt{Donnan.etal.2026} for spatially resolved analysis of the same PAHSPECS sample). The reconciliation of these disparate results within this unified framework further validates its robustness and highlights the fundamental role of the underlying physical conditions--especially the relative contributions of AGN irradiation and shock processing--in governing the manifestation and effectiveness of AGN feedback.

Beyond the two distinct trends of progressive PAH modifications associated with the two primary modes of AGN feedback, several higher-order variations also merit discussion. As discussed by \cite{Zhang.etal.2024b, Zhang.etal.2026}, shocks may also contribute to the production of smaller PAHs through the shattering and fragmentation of larger grains in highly obscured regions, where the resulting small PAHs can be effectively shielded from subsequent photodestruction (e.g., \citealt{Alonso-Herrero.etal.2014, Alonso-Herrero.etal.2020, Garcia-Bernete.etal.2022b, Garcia-Bernete.etal.2026}). This process helps explain the enhancement of both PAH 6.2/7.7 and 11.3/7.7 ratios observed in the nuclei of NGC~1097 and throughout NGC~4736, and likely also in the lower corners of panels (a5) and (b5) in Figures~\ref{PAHratioMaps}, which spatially coincide with the terminus of one of the spiraling gas arms feeding the circumnuclear disk in Circinus (e.g., \citealt{Izumi.etal.2018, Goesaert.etal.2025}). An alternative scenario, particularly for the (post-)starburst systems NGC~4736 and Circinus, is that the spaxels with larger PAH 6.2/7.7 ratios--indicating a higher fraction of small PAHs--reflect the bottom-up formation of PAHs from evolved stars. This interpretation is consistent with their comparatively enhanced PAH emission relative to the other three targets (see Figure~\ref{PAHvsSFR}), although the contribution of evolved stars to PAH production remains debated (for a review see \citealt{Li2020}). Nevertheless, the starburst nature of the nuclear region in Circinus, with $\Sigma_{\rm SFR}$ of $10\,M_{\odot}\,{\rm yr^{-1}}\,{\rm kpc^{-2}}$ (see Figure~\ref{PAHvsSFR} and also \citealt{For.etal.2012}), likely contributes to the ionization of PAHs, helping to explain the most extreme ionization fraction observed in this galaxy (see also \citealt{YuiDan.etal.2026} for a similar distribution of PAH band ratios among spatially resolved spaxels in an AGN--starburst composite). In particular, the $\rm rH_{2}$ values of most spaxels in Circinus are consistent with a significant contribution from star-formation activity to the excitation of H$_2$ (see Figure~\ref{H2NeP}), although ionized emission-line diagnostics indicate that all spaxels in Circinus are dominated by AGN excitation (see Figure~\ref{LineD}). We also note that NGC~1097 hosts a starburst ring at $r \approx 7-10\arcsec$, whereas the central $r \lesssim 2\arcsec$ region studied here is dominated by AGN activity (e.g., \citealt{Hummel.etal.1987, Hsieh.etal.2011, Mezcua&Prieto2014, Tabatabaei.etal.2018}), consistent with the relatively suppressed PAH emission observed in the nuclear region of this target (see Figure~\ref{PAHvsSFR}). Therefore, the enhanced PAH 6.2/7.7 and PAH 11.3/7.7 ratios observed in the nucleus of NGC~1097 are more likely explained by the former scenario, although we cannot entirely rule out the inward migration of newly formed PAHs from the starburst ring.

\subsection{Applications in Quantifying AGN Feedback}\label{sec4.3}

The above sections propose a unified framework for interpreting two distinct modes of AGN feedback operating in the central regions of AGN, highlighting the critical role of spatially resolved analysis for disentangling these feedback processes. In Section~\ref{sec3.1}, we discuss the practical application of our results in calibrating SFRs from PAH emission in AGN. In Section~\ref{sec3.2}, our findings highlight the strong potential of PAH features not only as tracers of SFRs, but also as quantitative diagnostics of AGN feedback, analogous to the role of traditional BPT diagrams in probing excitation mechanisms. Moreover, the results presented above suggest that AGN irradiation is preferentially associated with AGN ionization cones, whereas shock processing associated with either jets or outflows exhibits a broader spatial distribution and operates across a wider range of environments, including both LINERs and Seyfert galaxies. Developing this unified framework into a more rigorous quantitative one, including the establishment of anchor points for quantifying the relative contributions of different AGN feedback modes and the construction of a mixing-sequence PAH diagram to trace their interplay, will require extending the present analysis to a larger AGN sample that more comprehensively covers the full range of AGN luminosities and shock strengths.

Such a more rigorous quantitative framework, to be developed in future work, holds considerable promise for quantifying the relative impact of different AGN feedback modes in observations and thereby refining prescriptions for AGN-driven energy injection into the surrounding ISM in cosmological simulations. Specifically, this analysis has the potential to provide constraints for the structure and energetics of the AGN central engine in subgrid simulations, particularly when informed by observational constraints spanning a wide range of AGN luminosities and shock strengths. Theoretical studies indicate that AGN may operate in distinct accretion modes, from radiatively efficient, geometrically thin disks at high Eddington ratios to radiatively inefficient, geometrically thick flows at low accretion rates, which in turn regulate the balance between radiative and kinetic feedback (for reviews see \citealt{Yuan&Narayan2014, Heckman&Best2014}). These differences are expected to leave measurable imprints on the surrounding ISM, as observed in the PAH features, which can in turn provide constraints on the structure and energetics of the AGN central engine. In contrast to the more concentrated distribution of ionized emission-line ratios in the targets studied here (see Figure~\ref{LineD}), the broader and more continuous distribution of PAH band ratios suggests that PAH features more effectively trace the nonhomogeneous feedback processings associated with different AGN central engines, rather than the more general radiation-field strength traced by ionized emission lines.

JWST IFU spectroscopy also provides an opportunity to better constrain AGN duty cycles using tracers with complementary response timescales to nuclear activity. As discussed above, PAH molecules in AGN environments are vulnerable to destruction by feedback processings, and studies have estimated their lifetimes therein to be on the order of $\sim 10^{3}$–$10^{5}$ yr (\citealt{Voit1992, Micelotta.etal.2010b}). These timescales are intermediate between the near-instantaneous response of high-ionization coronal lines, and the longer dynamical timescales of large-scale multiphase outflows ($\gtrsim 10^{6}$ yr; for a review see \citealt{Veilleux.etal.2005}). Consequently, the spatially resolved distributions of PAH emission, coronal line strengths, and multiphase outflow kinematics--all accessible in the infrared--provide temporally sensitive probes of recent AGN activity, enabling the reconstruction of episodic accretion histories and tighter constraints on AGN lifetimes and duty cycles in subgrid prescriptions for cosmological simulations. These applications further demonstrate the potential of this unified framework for future studies to quantify the impact of AGN feedback and constrain the spatial and temporal scales over which different feedback modes operate.

\section{Summary}\label{sec5}

Using high-quality JWST MIRI/MRS IFU spectroscopy as part of the Galaxy Activity, Torus, and Outflow Survey (GATOS), this manuscript presents a spatially resolved analysis of diagnostic features from ionized gas, H$_2$, and PAHs in the central $r \approx 40-240$ pc regions of five carefully selected AGN. The sample consists of two Seyfert galaxies for which JWST spectroscopic observations are presented here for the first time, and three LINERs for which only nuclear integrated spectra have previously been analyzed. The observations probe physical scales of $\sim 4-24$ pc across the sample spanning nearly four orders of magnitude in AGN luminosity (log\,$L_{\rm bol}/{\rm erg\,s^{-1}} \approx 39.8 - 43.8$).

By combining measurements of the relative suppression of individual PAH features, the bimodal distributions of PAH band ratios, and tracers of shock processing and AGN irradiation, we identify two distinct AGN feedback modes that imprint characteristic signatures on PAH properties, associated with shock processing and AGN irradiation, respectively. These modes broadly correspond to the kinetic (or radio) and radiative (or quasar) forms of AGN feedback, underscoring the diagnostic power of PAH emission for probing AGN feedback processes. The spatially resolved analysis further reveals that the kinetic and radiative modes of AGN feedback can coexist in the central regions of AGN, although one mode generally dominates. Moreover, AGN irradiation appears to be preferentially associated with AGN ionization cones and is likely to become more important at higher redshifts. In contrast, shock processing--driven by either jets or outflows--exhibits a broader spatial distribution and is widespread among local AGN, including both LINERs and Seyfert galaxies.

The findings from the spatially resolved analysis provide new insights into calibrating SFRs from PAH emission in AGN and, more importantly, into distinguishing between different modes of AGN feedback. Together, the results and discussions presented here point to a practical unified framework for AGN feedback that incorporates both AGN irradiation and shock processing, enabling future investigations to quantify their relative contributions and constrain the spatial and temporal scales over which distinct feedback modes operate. This framework has the potential to provide critical observational benchmarks for modeling AGN-driven galaxy evolution, including constraints on the structure and energetics of the AGN central engine in subgrid simulations across a wide range of luminosities, as well as on AGN duty cycles through tracers with complementary response timescales to AGN activity.

\newpage

\acknowledgements
We thank the anonymous referee for the constructive and insightful comments, which have significantly improved the clarity and presentation of this manuscript. 

We thank Xiwen Zhu for assistance in preparing the schematic illustration presented in Figure~\ref{Illustration}. LZ and CP acknowledge grant support from the Space Telescope Science Institute (ID: JWST-GO-01670; JWST-GO-03535; JWST-GO-04225; JWST-GO-04972; JWST-GO-07429). EKSH and DD acknowledge support from the NASA Astrophysics Data Analysis Program (22-ADAP22-0173). OG-M acknowledge financial support from PAPIIT/DGAPA UNAM project IN109123, Ciencia de Frontera SECIHTI project CF-2023-G100, and sabbatical grant by PASPA/GDAPA at UNAM. RAR acknowledges the support from the Conselho Nacional de Desenvolvimento Cient\'{i}fico e Tecnol\'{o}gico (CNPq; Projects 303450/2022-3, and 403398/2023-1), the Coordena\c{c}\~{a}o de Aperfei\c{c}oamento de Pessoal de N\'{i}vel Superior (CAPES; Project 88887.894973/2023-00), and Funda\c{c}\~{a}o de Amparo \`{a} Pesquisa do Estado do Rio Grande do Sul (FAPERGS; Project 25/2551-0002765-9). MPS acknowledges support from grants RYC2021-033094-I, CNS2023-145506, and PID2023-146667NB-I00 funded by MCIN/AEI/10.13039/501100011033 and the European Union NextGenerationEU/PRTR. AJB acknowledges funding from the ``FirstGalaxies'' Advanced Grant from the European Research Council (ERC) under the European Union’s Horizon 2020 research and innovation program (Grant agreement No. 789056). EB cknowledge funding from the TEC-2024/TEC-182 project funded by the Comunidad de Madrid and from the grant PID2022-138621NB-I00, funded by MCIN/AEI/10.13039/501100011033/FEDER, EU. AA is funded by the European Union (Widening Participation, ExGal-Twin, GA 101158446.


\appendix

\section{Supplementary Figures}\label{secA0}
\begin{figure*}[!ht]
\figurenum{A1}
\center{\includegraphics[width=0.5\linewidth]{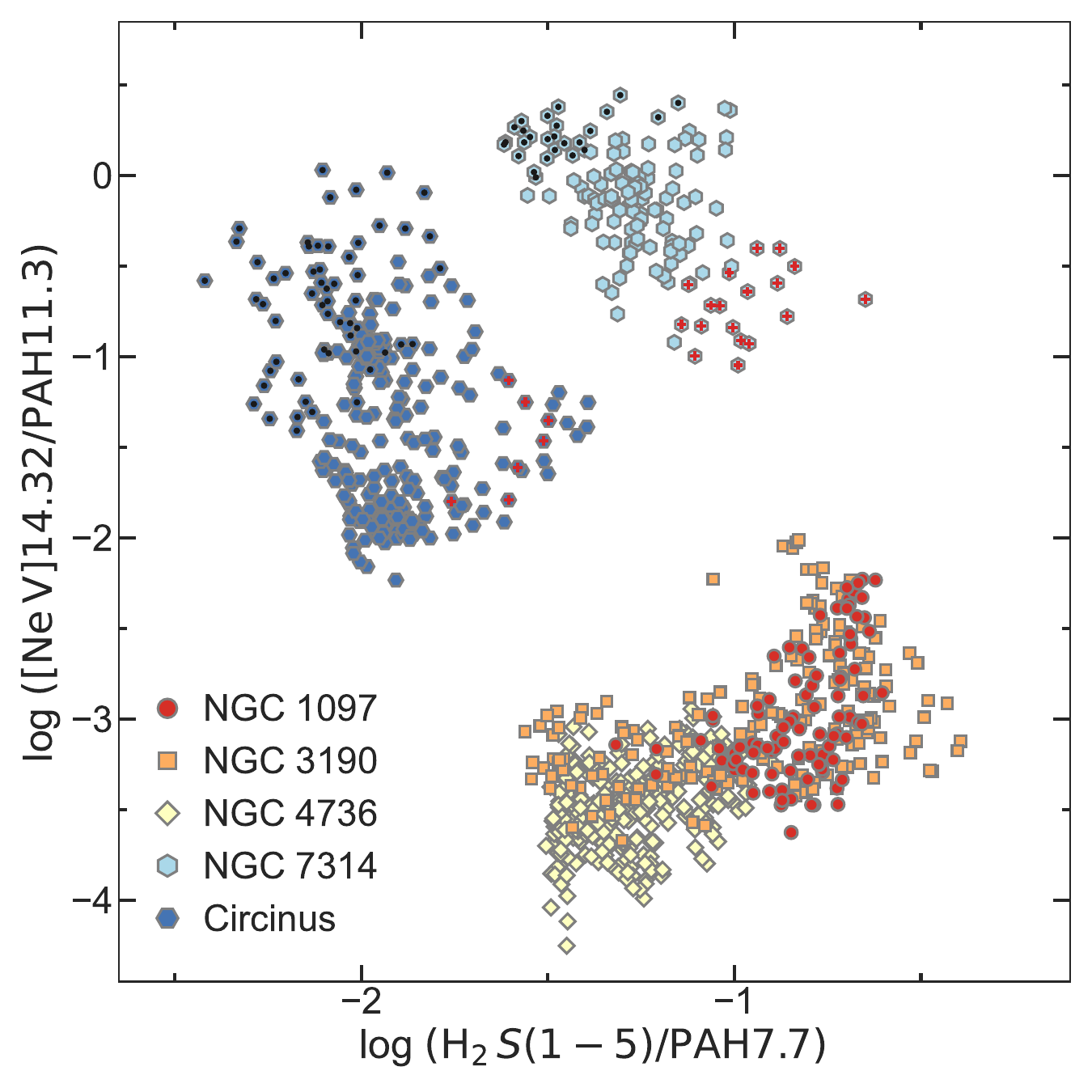}}
\caption{Diagnostic diagram of [Ne~{\small V}]14.32/PAH11.3 (i.e., $\rm rNe$) vs. H$_2$\,$S(1–5)$/PAH7.7 (i.e.,$\rm rH_{2}$) for $0\farcs2\times0\farcs2$ spaxels in the central $\sim 4\arcsec \times 4\arcsec$ regions of the five AGN targets. The data points highlighted with black dots pertain to spaxels in NGC~7314 and Circinus with PAH 11.3/7.7 $<0.2$ and PAH 6.2/7.7 $<0.15$, while the points highlighted with red crosses pertain to those with PAH 11.3/7.7 $>0.4$.}\label{H2NeP}
\end{figure*}

\begin{figure*}[!ht]
\figurenum{A2}
\center{\includegraphics[width=1\linewidth]{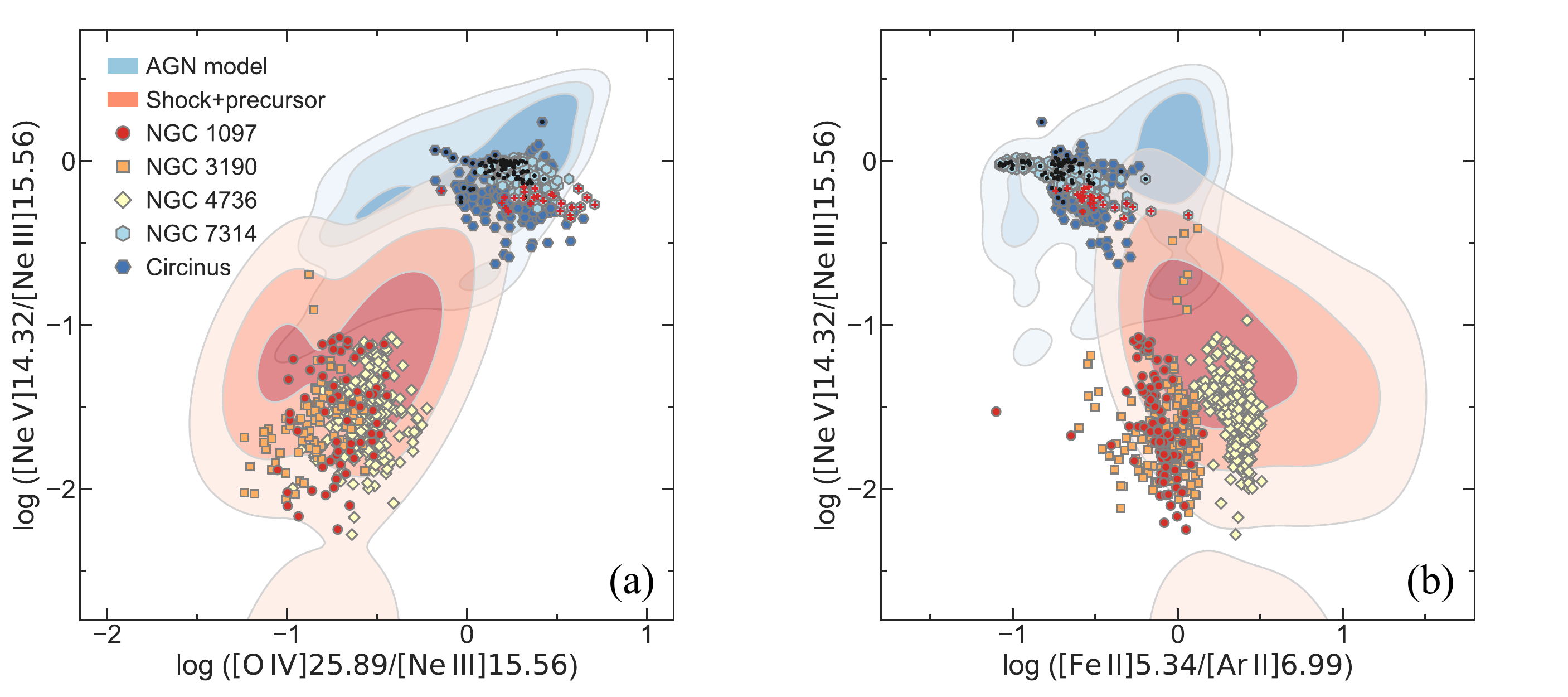}}
\caption{Diagnostic diagrams of ionized emission-line ratios: (a) [Ne~{\footnotesize V}]14.32/[Ne~{\footnotesize III}]15.56 vs. [O~{\footnotesize IV}]25.89/[Ne~{\footnotesize III}]15.56 and (b) [Ne~{\footnotesize V}]14.32/[Ne~{\footnotesize III}]15.56 vs. [Fe~{\footnotesize II}]5.34/[Ar~{\footnotesize II}]6.99 for $0\farcs2\times0\farcs2$ spaxels in the central $\sim 4\arcsec \times 4\arcsec$ regions of the five AGN targets. The data points with a black dot pertain to spaxels in NGC~7314 and Circinus with PAH 11.3/7.7 $<0.2$ and PAH 6.2/7.7 $<0.15$, while the points with a red cross pertain to those with PAH 11.3/7.7 $>0.4$. The bluish and reddish contours show the distributions of model results computed by \cite{Zhang.etal.2025} for AGN and fast radiative shocks (including the shock precursor), respectively. The included AGN models span ionization parameter log~$U$ from $-2.3$ to $-1.8$, AGN spectrum peak energy log $(E_{\rm peak}/{\rm keV})$ from $-2.1$ to $-0.9$, metallicities of $1.5-2\,Z_{\odot}$, and gas pressure log $(P/k)$ from $7.0$ to $9.5$. The included shock models cover shock velocities $v_s$ from $100 - 400\ \rm km\ s^{-1}$, pre-shock density $n_{\rm H}$ from $10^{3}$ to $10^{4}\ \rm cm^{-3}$, magnetic to ram pressure ratio $\eta_{\rm M}$ from $0.0$ to $0.001$, and metallicities of $1.5-2\,Z_{\odot}$. In each case, the contours enclose 30\%, 60\%, and 90\% of the model results, from the innermost to the outermost levels. Note that some spaxels have valid [Ne~{\footnotesize V}]14.32$\mu$m measurements from Ch3 but no corresponding [O~{\footnotesize IV}]25.89$\mu$m measurements from Ch4; consequently, panel (a) contains fewer data points than panel (b).}\label{LineD}
\end{figure*}

\begin{figure*}[!ht]
\figurenum{A3}
\center{\includegraphics[width=1\linewidth]{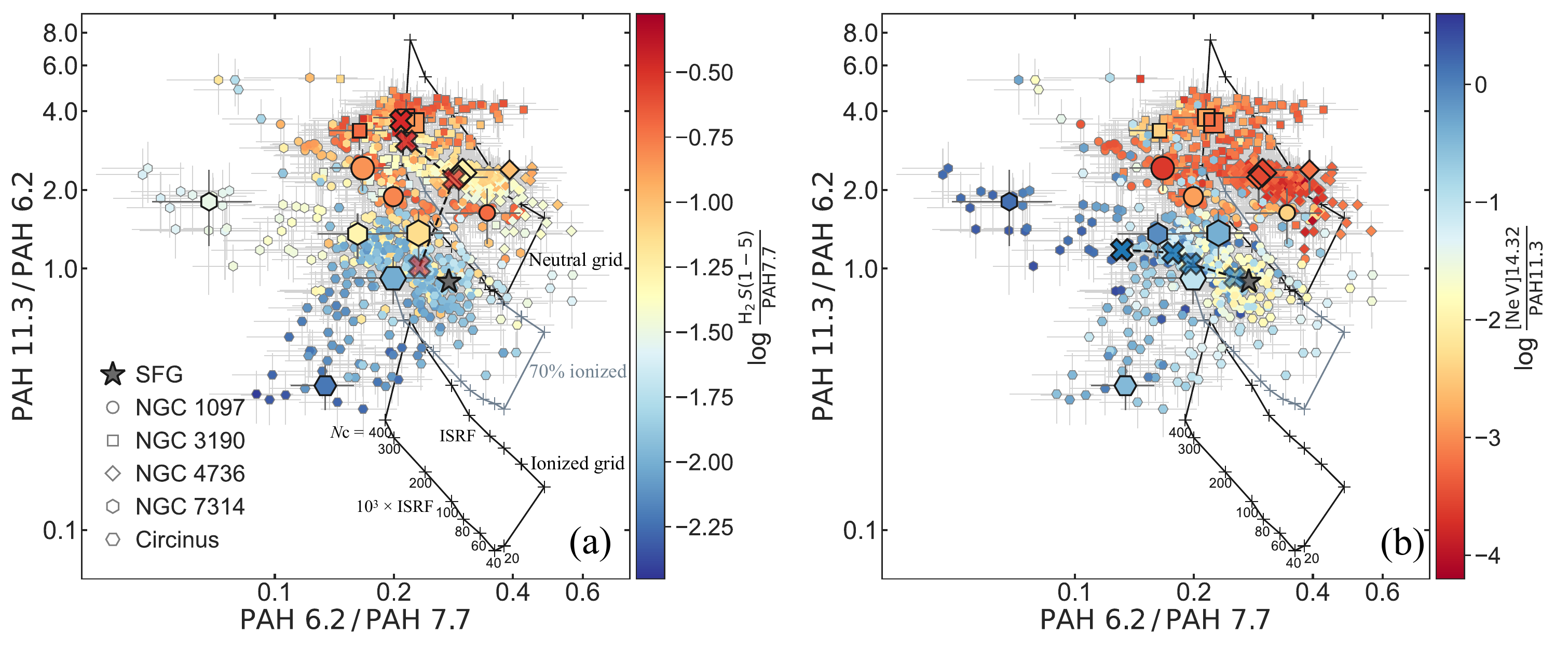}}
\caption{Same as Figure~\ref{PAHratios}, but with the y-axis replaced by the PAH~11.3/6.2 ratio. In addition, the PAH band ratios derived from the integrated spectra extracted within the $r \leq 0\farcs5$, $0\farcs5 \leq r \leq 1\farcs0$, and $1\farcs0 \leq r \leq 1\farcs5$ regions of each target are plotted with the same symbols and color coding as the spaxels in the five AGN, but are distinguished by black outlines of increasing size corresponding to the three radial apertures. Note that no data point is shown for the $r \leq 0\farcs5$ region of Circinus, as spaxels within this aperture are affected by saturation effects (see Figure~\ref{PAHratioMaps} and its caption).}\label{PAHratios_apx}
\end{figure*}

See Figures~\ref{H2NeP} and \ref{LineD} for the diagnostic diagram of [Ne~{\small V}]14.32/PAH11.3 vs H$_2$\,$S(1–5)$/PAH7.7 and the diagnostic diagrams of: (a) [Ne~{\small V}]/[Ne~{\small III}] vs. [O~{\small IV}]/[Ne~{\small III}] and (b) [Ne~{\small V}]/[Ne~{\small III}] vs. [Fe~{\small II}]/[Ar~{\small II}], respectively. See Figure~\ref{PAHratios_apx} for the PAH~11.3/6.2 versus PAH~6.2/7.7 diagnostic diagram, and see Figure~\ref{Fit_Demos} for examples illustrating the multi-component fitting procedure used to measure the PAH features. See Tables~\ref{tabcof1} and \ref{tabcof2} for the corresponding coefficients in Equations~\ref{equpp} and \ref{equss}, respectively. See Table~\ref{tabflux} for the flux measurements of the PAH 6.2, 7.7, and 11.3~$\mum$ features; the H$_2\,S(1)$--$S(6)$ transitions; and the [Ne~{\footnotesize II}]12.81$\mu$m, [Ne~{\footnotesize III}]15.555$\mu$m, [Ne~{\footnotesize V}]14.32$\mu$m, [Fe~{\footnotesize II}]5.34$\mu$m, and [Ar~{\footnotesize II}]6.985$\mu$m emission lines measured from the aperture-extracted spectra shown in Figure~\ref{Fit_Demos}.


\startlongtable
\tablenum{A1}
\setlength{\tabcolsep}{3pt}
\begin{deluxetable*}{cccccc}
\tabletypesize{\footnotesize}
\tablecolumns{6}
\tablecaption{Coefficients in Equation~\ref{equpp}}
\tablehead{
\colhead{Feature} & \colhead{$\alpha_{1}$} & \colhead{$\alpha_{2}$} & \colhead{$\alpha_{3}$} & \colhead{$\beta$} & \colhead{$\sigma$ (dex)}\\
\colhead{(1)} & \colhead{(2)} & \colhead{(3)} & \colhead{(4)} & \colhead{(5)} & \colhead{(6)} }
\startdata
PAH6.2 & 0.87$\pm$0.02& $-$0.90$\pm$0.03 & $-$0.26$\pm$0.01 & 38.68$\pm$0.12 & 0.25 \\
PAH7.7 & 0.88$\pm$0.01& $-$0.86$\pm$0.02 & $-$0.19$\pm$0.01 & 39.60$\pm$0.10 & 0.19 \\
PAH11.3 & 0.78$\pm$0.02& $-$0.64$\pm$0.02 & $-$0.32$\pm$0.01 & 38.59$\pm$0.11 & 0.21 \\
\enddata
\label{tabcof1}
\end{deluxetable*}

\startlongtable
\tablenum{A2}
\setlength{\tabcolsep}{3pt}
\begin{deluxetable*}{cccccc}
\tabletypesize{\footnotesize}
\tablecolumns{6}
\tablecaption{Coefficients in Equation~\ref{equss}}
\tablehead{
\colhead{Feature} & \colhead{$\alpha_{1}$} & \colhead{$\alpha_{2}$} & \colhead{$\alpha_{3}$} & \colhead{$\beta$} & \colhead{$\sigma$ (dex)}\\
\colhead{(1)} & \colhead{(2)} & \colhead{(3)} & \colhead{(4)} & \colhead{(5)} & \colhead{(6)} }
\startdata
PAH6.2 & 0.83$\pm$0.02& 0.50$\pm$0.03 & 0.25$\pm$0.01 & $-$34.05$\pm$0.54 & 0.24 \\
PAH7.7 & 0.93$\pm$0.01& 0.63$\pm$0.03 & 0.20$\pm$0.01 & $-$38.04$\pm$0.47 & 0.20 \\
PAH11.3 & 0.94$\pm$0.02& 0.38$\pm$0.03 & 0.34$\pm$0.01 & $-$38.25$\pm$0.60 & 0.24 \\
\enddata
\label{tabcof2}
\end{deluxetable*}

\startlongtable
\tablenum{A3}
\setlength{\tabcolsep}{1pt}
\begin{deluxetable*}{cccccccccccccccc}
\tabletypesize{\tiny}
\tablecolumns{16}
\tablecaption{Flux Measurements from Aperture-extracted Spectra}
\tablehead{
\colhead{Aperture} & \colhead{log\,$f_{\rm PAH}^{6.2}$} & \colhead{log\,$f_{\rm PAH}^{7.7}$} & \colhead{log\,$f_{\rm PAH}^{11.3}$} & \colhead{log\,$f_{\rm H_2}^{S(1)}$} & \colhead{log\,$f_{\rm H_2}^{S(2)}$} & \colhead{log\,$f_{\rm H_2}^{S(3)}$} & \colhead{log\,$f_{\rm H_2}^{S(4)}$} & \colhead{log\,$f_{\rm H_2}^{S(5)}$} & \colhead{log\,$f_{\rm H_2}^{S(6)}$} & \colhead{log\,$f_{\rm [Ne\,II]}^{12.81}$} & \colhead{log\,$f_{\rm [Ne\,III]}^{15.56}$} & \colhead{log\,$f_{\rm [Ne\,V]}^{14.32}$} & \colhead{log\,$f_{\rm [Fe\,II]}^{5.34}$} & \colhead{log\,$f_{\rm [Ar\,II]}^{6.99}$} & \colhead{log\,$\tau_{9.7}$} \\
\colhead{(1)} & \colhead{(2)} & \colhead{(3)} & \colhead{(4)} & \colhead{(5)} & \colhead{(6)} & \colhead{(7)} & \colhead{(8)} & \colhead{(9)} & \colhead{(10)} & \colhead{(11)} & \colhead{(12)} & \colhead{(13)} & \colhead{(14)} & \colhead{(15)} & \colhead{(16)} }
\startdata
 NGC1097\_0.5 & $-$13.20$\pm$0.07 & $-$12.74$\pm$0.06 & $-$12.99$\pm$0.08 & $-$14.08$\pm$0.01 & $-$14.21$\pm$0.01 & $-$14.03$\pm$0.01 & $-$14.34$\pm$0.01 & $-$14.09$\pm$0.01 & $-$14.77$\pm$0.01 & $-$13.90$\pm$0.04 & $-$14.21$\pm$0.03 & $-$15.39$\pm$0.03 & $-$14.58$\pm$0.01 & $-$14.34$\pm$0.05 & $-$9.99 \\
 NGC1097\_1.0 & $-$13.14$\pm$0.03 & $-$12.44$\pm$0.03 & $-$12.86$\pm$0.04 & $-$13.81$\pm$0.01 & $-$13.99$\pm$0.01 & $-$13.84$\pm$0.01 & $-$14.19$\pm$0.01 & $-$13.95$\pm$0.01 & $-$14.61$\pm$0.01 & $-$13.90$\pm$0.01 & $-$14.21$\pm$0.02 & $-$15.73$\pm$0.11 & $-$14.52$\pm$0.01 & $-$14.40$\pm$0.03 & $-$8.49 \\
 NGC1097\_1.5 & $-$13.26$\pm$0.06 & $-$12.48$\pm$0.05 & $-$12.87$\pm$0.05 & $-$13.81$\pm$0.02 & $-$14.07$\pm$0.02 & $-$13.92$\pm$0.01 & $-$14.34$\pm$0.01 & $-$14.10$\pm$0.01 & $-$14.75$\pm$0.01 & $-$13.97$\pm$0.18 & $-$14.28$\pm$0.01 & $-$16.51$\pm$0.08 & $-$14.68$\pm$0.01 & $-$14.59$\pm$0.12 & $-$9.99 \\
 NGC3190\_0.5 & $-$13.79$\pm$0.05 & $-$13.00$\pm$0.05 & $-$13.26$\pm$0.04 & $-$14.50$\pm$0.01 & $-$14.61$\pm$0.01 & $-$14.22$\pm$0.01 & $-$14.58$\pm$0.01 & $-$14.25$\pm$0.01 & $-$14.91$\pm$0.01 & $-$14.11$\pm$0.03 & $-$14.25$\pm$0.01 & $-$15.67$\pm$0.01 & $-$14.55$\pm$0.01 & $-$14.53$\pm$0.03 & $-$0.88 \\
 NGC3190\_1.0 & $-$13.51$\pm$0.06 & $-$12.85$\pm$0.05 & $-$12.94$\pm$0.05 & $-$14.22$\pm$0.01 & $-$14.35$\pm$0.01 & $-$14.00$\pm$0.01 & $-$14.35$\pm$0.02 & $-$14.02$\pm$0.01 & $-$14.71$\pm$0.02 & $-$13.93$\pm$0.06 & $-$14.07$\pm$0.02 & $-$15.74$\pm$0.01 & $-$14.32$\pm$0.01 & $-$14.36$\pm$0.03 & $-$7.98 \\
 NGC3190\_1.5 & $-$13.34$\pm$0.03 & $-$12.69$\pm$0.02 & $-$12.78$\pm$0.01 & $-$14.15$\pm$0.02 & $-$14.39$\pm$0.03 & $-$14.09$\pm$0.03 & $-$14.49$\pm$0.03 & $-$14.17$\pm$0.03 & $-$14.96$\pm$0.04 & $-$13.98$\pm$0.09 & $-$14.14$\pm$0.08 & $-$16.00$\pm$0.03 & $-$14.40$\pm$0.01 & $-$14.41$\pm$0.05 & $-$7.88 \\
 NGC4736\_0.5 & $-$13.05$\pm$0.06 & $-$12.65$\pm$0.05 & $-$12.68$\pm$0.05 & $-$14.36$\pm$0.01 & $-$14.54$\pm$0.01 & $-$14.15$\pm$0.01 & $-$14.58$\pm$0.01 & $-$14.22$\pm$0.01 & $-$14.91$\pm$0.04 & $-$14.38$\pm$0.01 & $-$14.41$\pm$0.03 & $-$15.89$\pm$0.02 & $-$14.31$\pm$0.01 & $-$14.60$\pm$0.05 & $-$7.99 \\
 NGC4736\_1.0 & $-$12.67$\pm$0.06 & $-$12.13$\pm$0.05 & $-$12.32$\pm$0.06 & $-$13.98$\pm$0.01 & $-$14.21$\pm$0.02 & $-$13.88$\pm$0.01 & $-$14.36$\pm$0.02 & $-$14.03$\pm$0.01 & $-$14.76$\pm$0.04 & $-$14.17$\pm$0.01 & $-$14.17$\pm$0.08 & $-$15.64$\pm$0.01 & $-$14.08$\pm$0.01 & $-$14.38$\pm$0.04 & $-$7.98 \\
 NGC4736\_1.5 & $-$12.49$\pm$0.03 & $-$11.97$\pm$0.02 & $-$12.13$\pm$0.03 & $-$13.77$\pm$0.02 & $-$14.06$\pm$0.02 & $-$13.76$\pm$0.02 & $-$14.29$\pm$0.01 & $-$13.96$\pm$0.01 & $-$14.71$\pm$0.03 & $-$14.09$\pm$0.01 & $-$14.09$\pm$0.04 & $-$15.66$\pm$0.01 & $-$14.00$\pm$0.01 & $-$14.30$\pm$0.04 & $-$7.90 \\
 NGC7314\_0.5 & $-$13.41$\pm$0.12 & $-$12.24$\pm$0.04 & $-$13.15$\pm$0.08 & $-$14.38$\pm$0.03 & $-$14.62$\pm$0.03 & $-$14.37$\pm$0.04 & $-$14.66$\pm$0.02 & $-$14.37$\pm$0.02 & $-$15.47$\pm$0.21 & $-$13.37$\pm$0.10 & $-$12.97$\pm$0.03 & $-$12.99$\pm$0.07 & $-$14.66$\pm$0.01 & $-$13.68$\pm$0.01 & $-$0.60 \\
 NGC7314\_1.0 & $-$13.27$\pm$0.06 & $-$12.48$\pm$0.06 & $-$13.14$\pm$0.06 & $-$14.27$\pm$0.02 & $-$14.58$\pm$0.02 & $-$14.35$\pm$0.03 & $-$14.79$\pm$0.02 & $-$14.52$\pm$0.02 & $-$15.34$\pm$0.08 & $-$13.57$\pm$0.01 & $-$13.18$\pm$0.01 & $-$13.21$\pm$0.04 & $-$14.73$\pm$0.01 & $-$13.92$\pm$0.01 & $-$0.50 \\
 NGC7314\_1.5 & $-$13.55$\pm$0.06 & $-$12.92$\pm$0.06 & $-$13.42$\pm$0.06 & $-$14.45$\pm$0.02 & $-$14.85$\pm$0.01 & $-$14.66$\pm$0.02 & $-$15.22$\pm$0.02 & $-$14.97$\pm$0.02 & $-$15.64$\pm$0.02 & $-$14.14$\pm$0.06 & $-$13.65$\pm$0.02 & $-$13.80$\pm$0.03 & $-$15.13$\pm$0.01 & $-$14.54$\pm$0.01 & $-$0.71 \\
 Circinus\_1.0 & $-$11.64$\pm$0.06 & $-$10.77$\pm$0.05 & $-$12.09$\pm$0.06 & $-$13.74$\pm$0.02 & $-$13.88$\pm$0.02 & $-$13.57$\pm$0.01 & $-$13.83$\pm$0.01 & $-$13.51$\pm$0.01 & $-$14.32$\pm$0.08 & $-$12.50$\pm$0.01 & $-$12.55$\pm$0.03 & $-$12.59$\pm$0.08 & $-$13.55$\pm$0.01 & $-$12.81$\pm$0.01 & 0.09 \\
 Circinus\_1.5 & $-$11.24$\pm$0.07 & $-$10.53$\pm$0.06 & $-$11.27$\pm$0.06 & $-$13.25$\pm$0.02 & $-$13.35$\pm$0.02 & $-$13.16$\pm$0.02 & $-$13.41$\pm$0.02 & $-$13.11$\pm$0.02 & $-$13.98$\pm$0.11 & $-$12.22$\pm$0.04 & $-$12.20$\pm$0.01 & $-$12.29$\pm$0.02 & $-$13.19$\pm$0.01 & $-$12.54$\pm$0.01 & $-$0.17 \\
\enddata
\tablecomments{\footnotesize Flux measurements from integrated MIRI/MRS spectra extracted from the $r \leq 0\farcs5$, $0\farcs5 \leq r \leq 1\farcs0$, and $1\farcs0 \leq r \leq 1\farcs5$ regions of each target. All measurements have been corrected for dust extinction based on the fitted $\tau_{9.7}$ values listed in the last column (see Section~\ref{sec2.3}).}
\label{tabflux}
\end{deluxetable*}

\begin{figure*}[!ht]
\figurenum{A4}
\center{\includegraphics[width=0.85\linewidth]{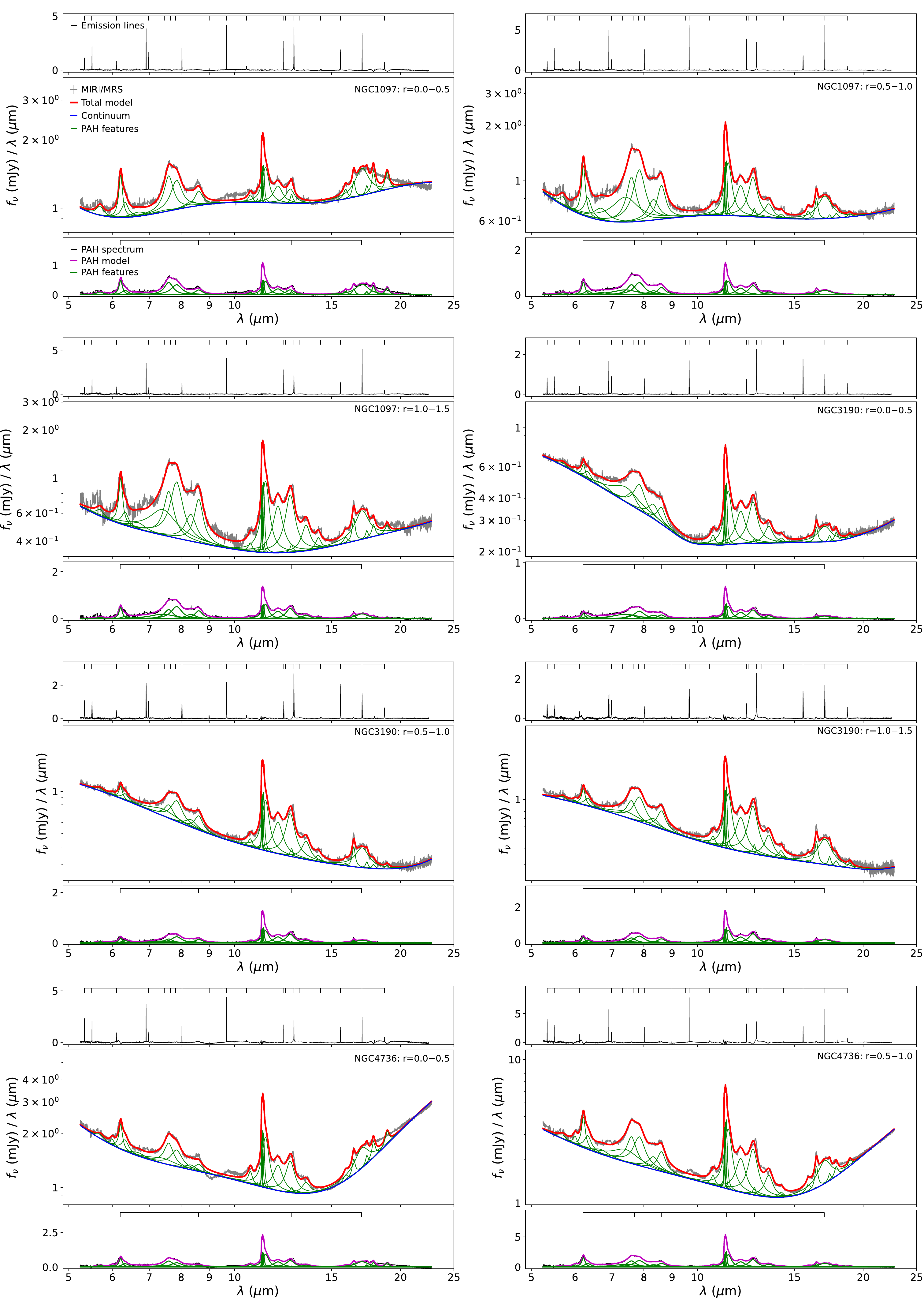}}
\end{figure*}

\begin{figure*}[!ht]
\figurenum{A4}
\center{\includegraphics[width=0.85\linewidth]{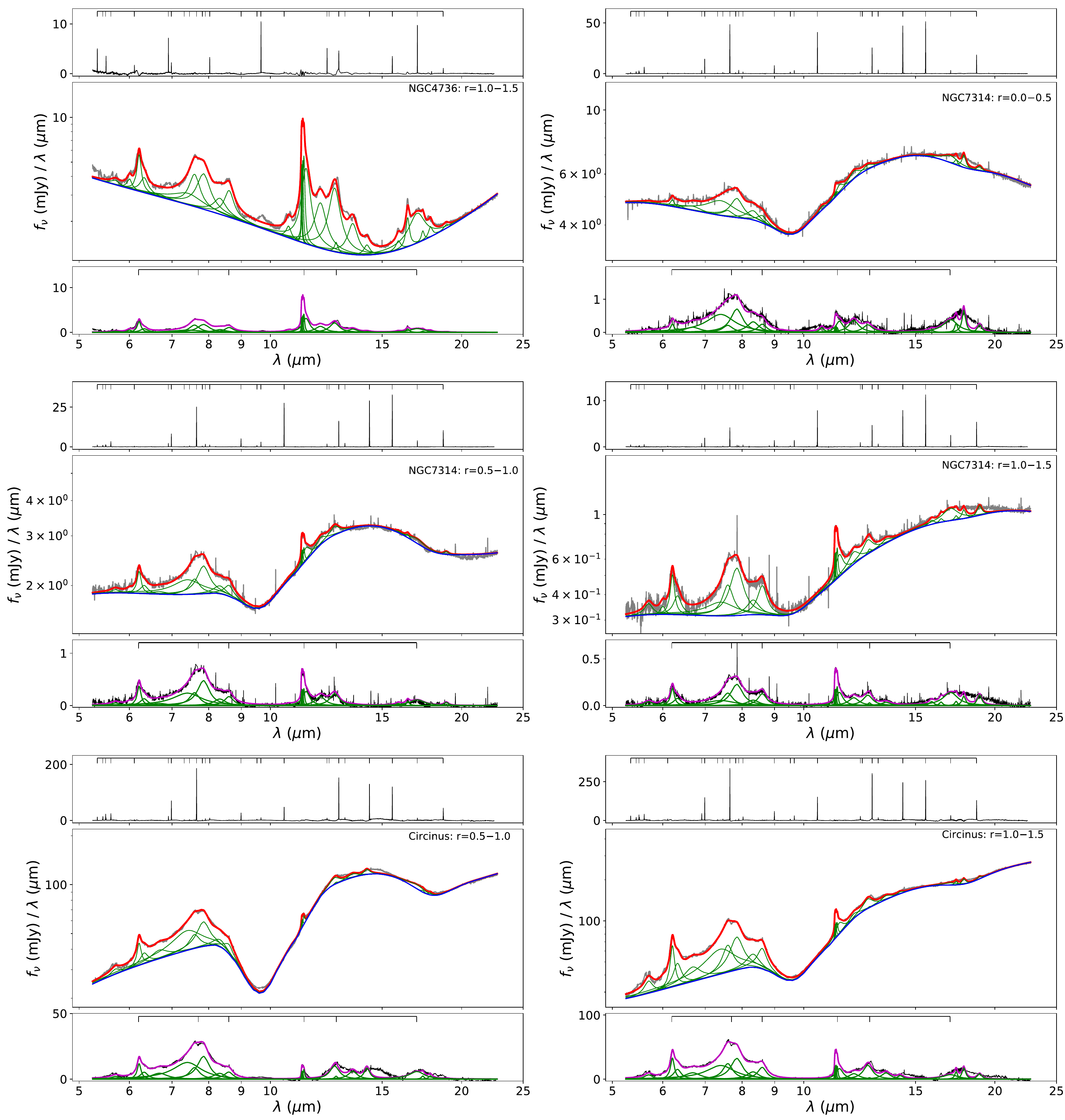}}
\caption{Top panels: Emission-line spectra obtained by subtracting the PAH emission and underlying continuum (i.e., the red curves in the middle panels) from the integrated MIRI/MRS spectra extracted from the $r \leq 0\farcs5$, $0\farcs5 \leq r \leq 1\farcs0$, and $1\farcs0 \leq r \leq 1\farcs5$ regions of each target. From left to right, the short markers at the top indicate the positions of the [Fe~{\footnotesize II}]5.34$\mu$m, [Fe~{\footnotesize VIII}]5.45$\mu$m, [Mg~{\footnotesize VII}]5.50$\mu$m, [Mg~{\footnotesize V}]5.61$\mu$m, ${\rm H}_{2}\, S(6)$, ${\rm H}_{2}\, S(5)$, [Ar~{\footnotesize II}]6.985$\mu$m, [Na~{\footnotesize III}]7.32$\mu$m, Pf$\alpha$, [Ne~{\footnotesize VI}]7.65$\mu$m, [Fe~{\footnotesize VII}]7.815$\mu$m, [Ar~{\footnotesize V}]7.90$\mu$m, ${\rm H}_{2}\, S(4)$, [Ar~{\footnotesize III}]8.99$\mu$m, [Fe~{\footnotesize VII}]9.53$\mu$m, ${\rm H}_{2}\, S(3)$, [S~{\footnotesize IV}]10.51$\mu$m, ${\rm H}_{2}\, S(2)$, Hu$\alpha$, [Ne~{\footnotesize II}]12.81$\mu$m, [Ar~{\footnotesize V}]13.10$\mu$m, [Ne~{\footnotesize V}]14.32$\mu$m, [Ne~{\footnotesize III}]15.555$\mu$m, ${\rm H}_{2}\, S(1)$, and [S~{\footnotesize III}]18.71$\mu$m emission lines, respectively. Middle panels: Illustrations of the multi-component fitting procedure used for the PAH measurements (see Section~\ref{sec2.3}). The gray curves show the observed MIRI/MRS spectra with emission lines masked. The red curves represent the best-fit models consisting of PAH emission and the underlying continuum. The blue curves show the combined underlying continuum components (i.e., stellar and dust continua), while the green curves represent the individual PAH features modeled with Drude profiles. Bottom panels: PAH spectra (black curves) obtained by subtracting all continuum components (i.e., the blue curves in the middle panels) from the emission-line-masked MIRI/MRS spectra. As in the middle panels, the green curves represent individual PAH features modeled with Drude profiles, while the magenta curves show their sum, corresponding to the modeled PAH spectra. From left to right, the short markers at the top indicate the positions of the PAH complexes centered near 6.2, 7.7, 8.6, 11.3, 12.7, and 17.0~$\mum$, respectively. Note that all x-axes, as well as the y-axes in the middle panels, are shown on logarithmic scales, whereas the y-axes in the top and bottom panels are shown on linear scales. No spectrum is shown for the $r \leq 0\farcs5$ region of Circinus, as noted in Figure~\ref{PAHratios_apx}.}\label{Fit_Demos}
\end{figure*}


\begin{thebibliography}{}
\expandafter\ifx\csname natexlab\endcsname\relax\def\natexlab#1{#1}\fi

\bibitem[Aitken \& Roche(1985)]{Aitken&Roche1985} Aitken, D.~K. \& Roche, P.~F.\ 1985, \mnras, 213, 777

\bibitem[Alexander \& Hickox(2012)]{Alexander&Hickox2012} Alexander, D.~M. \& Hickox, R.~C.\ 2012, \nar, 56, 4, 93

\bibitem[Allain et al.(1996)]{Allain.etal.1996} Allain, T., Leach, S., \& Sedlmayr, E.\ 1996, \aap, 305, 616

\bibitem[Allen et al.(2008)]{Allen.etal.2008} Allen, M.~G., Groves, B.~A., Dopita, M.~A., et al.\ 2008, \apjs, 178, 1, 20

\bibitem[Alonso-Herrero et al.(2021)]{AlonsoHerrero.etal.2021} Alonso-Herrero, A., Garc{\'\i}a-Burillo, S., H{\"o}nig, S.~F., et al.\ 2021, \aap, 652, A99

\bibitem[Alonso Herrero et al.(2025)]{AlonsoHerrero.etal.2025} Alonso Herrero, A., Hermosa Mu{\~n}oz, L., Labiano, A., et al.\ 2025, \aap, 699, A334

\bibitem[Alonso-Herrero et al.(2020)]{Alonso-Herrero.etal.2020} Alonso-Herrero, A., Pereira-Santaella, M., Rigopoulou, D., et al.\ 2020, \aap, 639, A43

\bibitem[Alonso-Herrero et al.(2014)]{Alonso-Herrero.etal.2014} Alonso-Herrero, A., Ramos Almeida, C., Esquej, P., et al.\ 2014, \mnras, 443, 2766

\bibitem[Annuar et al.(2025)]{Annuar.etal.2025} Annuar, A., Alexander, D.~M., Gandhi, P., et al.\ 2025, \mnras, 540, 4, 3827

\bibitem[Argyriou et al.(2023)]{Argyriou.etal.2023} Argyriou, I., Glasse, A., Law, D.~R., et al.\ 2023, \aap, 675, A111

\bibitem[Audibert et al.(2023)]{Audibert.etal.2023} Audibert, A., Ramos Almeida, C., Garc{\'\i}a-Burillo, S., et al.\ 2023, \aap, 671, L12

\bibitem[Bellocchi et al.(2026)]{Bellocchi.etal.2026} Bellocchi, E., Longinotti, A.~L., Salom{\'e}, Q., et al.\ 2026, \aap, 709, A260

\bibitem[Bierschenk et al.(2024)]{Bierschenk.etal.2024} Bierschenk, M., Ricci, C., Temple, M.~J., et al.\ 2024, \apj, 976, 2, 257

\bibitem[Bushouse et al.(2025)]{Bushouse.etal.2025} Bushouse, H., Eisenhamer, J., Dencheva, N., et al.\ 2025, Zenodo, 1.18.0. doi:10.5281/zenodo.15178003

\bibitem[Chown et al.(2025)]{Chown.etal.2025} Chown, R., Okada, Y., Peeters, E., et al.\ 2025, \aap, 698, A86.

\bibitem[Cicone et al.(2014)]{Cicone.etal.2014} Cicone, C., Maiolino, R., Sturm, E., et al.\ 2014, \aap, 562, A21

\bibitem[Cisternas et al.(2013)]{Cisternas.etal.2013} Cisternas, M., Gadotti, D.~A., Knapen, J.~H., et al.\ 2013, \apj, 776, 1, 50

\bibitem[Colina et al.(2015)]{Colina.etal.2015} Colina, L., Piqueras L{\'o}pez, J., Arribas, S., et al.\ 2015, \aap, 578, A48

\bibitem[Cortzen et al.(2019)]{Cortzen.etal.2019} Cortzen, I., Garrett, J., Magdis, G., et al.\ 2019, \mnras, 482, 2, 1618

\bibitem[Costa-Souza et al.(2026)]{Costa-Souza.etal.2026} Costa-Souza, J.~H., Colina, L., Riffel, R.~A., et al.\ 2026, arXiv:2605.04925

\bibitem[Crain et al.(2015)]{Crain.etal.2015} Crain, R.~A., Schaye, J., Bower, R.~G., et al.\ 2015, \mnras, 450, 2, 1937

\bibitem[da Silva et al.(2023)]{daSilva.etal.2023} da Silva, P., Menezes, R.~B., D{\'\i}az, Y., et al.\ 2023, \mnras, 519, 1, 1293

\bibitem[Dasyra et al.(2024)]{Dasyra.etal.2024} Dasyra, K.~M., Paraschos, G.~F., Combes, F., et al.\ 2024, \apj, 977, 2, 156

\bibitem[Dav{\'e} et al.(2020)]{Dave.etal.2020} Dav{\'e}, R., Crain, R.~A., Stevens, A.~R.~H., et al.\ 2020, \mnras, 497, 1, 146

\bibitem[Davies et al.(2014)]{Davies.etal.2014} Davies, R.~I., Maciejewski, W., Hicks, E.~K.~S., et al.\ 2014, \apj, 792, 2, 101

\bibitem[Davies et al.(2024)]{Davies.etal.2024} Davies, R., Shimizu, T., Pereira-Santaella, M., et al.\ 2024, \aap, 689, A263

\bibitem[Delaney et al.(2026)]{Delaney.etal.2026} Delaney, D.~E., Hicks, E.~K.~S., Zhang, L., et al.\ 2026, \apj, 1002, 1, 20

\bibitem[Di Matteo et al.(2005)]{DiMatteo.etal.2005} Di Matteo, T., Springel, V., \& Hernquist, L.\ 2005, \nat, 433, 7026, 604

\bibitem[Diamond-Stanic \& Rieke(2010)]{Diamond-Stanic&Rieke2010} Diamond-Stanic, A.~M. \& Rieke, G.~H.\ 2010, \apj, 724, 140

\bibitem[Dors et al.(2012)]{Dors.etal.2012} Dors, O.~L., Riffel, R.~A., Cardaci, M.~V., et al.\ 2012, \mnras, 422, 1, 252

\bibitem[Donnan et al.(2024)]{Donnan.etal.2024} Donnan, F.~R., Garc{\'\i}a-Bernete, I., Rigopoulou, D., et al.\ 2024, \mnras, 529, 2, 1386

\bibitem[Donnan et al.(2026)]{Donnan.etal.2026} Donnan, F.~R., Sandstrom, K., Shivaei, I., et al.\ 2026, \mnras, in press, arXiv:2606.18244

\bibitem[Dopita et al.(2015)]{Dopita.etal.2015} Dopita, M.~A., Ho, I.-T., Dressel, L.~L., et al.\ 2015, \apj, 801, 1, 42

\bibitem[Draine \& Li(2007)]{Draine&Li2007} Draine, B.~T. \& Li, A.\ 2007, \apj, 657, 810

\bibitem[Draine et al.(2021)]{Draine.etal.2021} Draine, B.~T., Li, A., Hensley, B.~S., et al.\ 2021, \apj, 917, 3

\bibitem[Durr{\'e} \& Mould(2018)]{Durre&Mould2018} Durr{\'e}, M. \& Mould, J.\ 2018, \apj, 867, 2, 149

\bibitem[Greene et al.(2020)]{Greene.etal.2020} Greene, J.~E., Strader, J., \& Ho, L.~C.\ 2020, \araa, 58, 257

\bibitem[Esparza-Arredondo et al.(2018)]{Esparza-Arredondo.etal.2018} Esparza-Arredondo, D., Gonz{\'a}lez-Mart{\'\i}n, O., Dultzin, D., et al.\ 2018, \apj, 859, 2, 124

\bibitem[Esposito et al.(2024)]{Esposito.etal.2024} Esposito, F., Alonso-Herrero, A., Garc{\'\i}a-Burillo, S., et al.\ 2024, \aap, 686, A46

\bibitem[Esquej et al.(2014)]{Esquej.etal.2014} Esquej, P., Alonso-Herrero, A., Gonz{\'a}lez-Mart{\'\i}n, O., et al.\ 2014, \apj, 780, 1, 86

\bibitem[Fabian(2012)]{Fabian2012} Fabian, A.~C.\ 2012, \araa, 50, 455

\bibitem[Fern{\'a}ndez-Ontiveros et al.(2012)]{Fernandez-Ontiveros.etal.2012} Fern{\'a}ndez-Ontiveros, J.~A., Prieto, M.~A., Acosta-Pulido, J.~A., et al.\ 2012, Journal of Physics Conference Series, 372, 1, 012006

\bibitem[Fiore et al.(2017)]{Fiore.etal.2017} Fiore, F., Feruglio, C., Shankar, F., et al.\ 2017, \aap, 601, A143

\bibitem[Fluetsch et al.(2019)]{Fluetsch.etal.2019} Fluetsch, A., Maiolino, R., Carniani, S., et al.\ 2019, \mnras, 483, 4, 4586

\bibitem[For et al.(2012)]{For.etal.2012} For, B.-Q., Koribalski, B.~S., \& Jarrett, T.~H.\ 2012, \mnras, 425, 3, 1934

\bibitem[Forbes \& Ward(1993)]{Forbes&Ward1993} Forbes, D.~A. \& Ward, M.~J.\ 1993, \apj, 416, 150

\bibitem[Foreman-Mackey et al.(2013)]{Foreman-Mackey.etal.2013} Foreman-Mackey, D., Hogg, D.~W., Lang, D., et al.\ 2013, \pasp, 125, 925, 306

\bibitem[Garc{\'\i}a-Bernete et al.(2026)]{Garcia-Bernete.etal.2026} Garc{\'\i}a-Bernete, I., Pereira-Santaella, M., Gonz{\'a}lez-Alfonso, E., et al.\ 2026, Nature Astronomy, 10, 420

\bibitem[Garc{\'\i}a-Bernete et al.(2015)]{Garcia-Bernete.etal.2015} Garc{\'\i}a-Bernete, I., Ramos Almeida, C., Acosta-Pulido, J.~A., et al.\ 2015, \mnras, 449, 2, 1309

\bibitem[Garc{\'\i}a-Bernete et al.(2017)]{Garcia-Bernete.etal.2017} Garc{\'\i}a-Bernete, I., Ramos Almeida, C., Landt, H., et al.\ 2017, \mnras, 469, 1, 110

\bibitem[Garc{\'\i}a-Bernete et al.(2022a)]{Garcia-Bernete.etal.2022a} Garc{\'\i}a-Bernete, I., Rigopoulou, D., Alonso-Herrero, A., et al.\ 2022a, \aap, 666, L5

\bibitem[Garc{\'\i}a-Bernete et al.(2022b)]{Garcia-Bernete.etal.2022b} Garc{\'\i}a-Bernete, I., Rigopoulou, D., Alonso-Herrero, A., et al.\ 2022b, \mnras, 509, 4256

\bibitem[Garc{\'\i}a-Bernete et al.(2024b)]{Garcia-Bernete.etal.2024b} Garc{\'\i}a-Bernete, I., Rigopoulou, D., Donnan, F.~R., et al.\ 2024b, \aap, 691, A162

\bibitem[Garc{\'\i}a-Burillo et al.(2021)]{Garcia-Burillo.etal.2021} Garc{\'\i}a-Burillo, S., Alonso-Herrero, A., Ramos Almeida, C., et al.\ 2021, \aap, 652, A98

\bibitem[Garc{\'\i}a-Burillo et al.(2024)]{Garcia-Burillo.etal.2024} Garc{\'\i}a-Burillo, S., Hicks, E.~K.~S., Alonso-Herrero, A., et al.\ 2024, \aap, 689, A347

\bibitem[Gardner et al.(2023)]{Gardner.etal.2023} Gardner, J.~P., Mather, J.~C., Abbott, R., et al.\ 2023, \pasp, 135, 1048, 068001

\bibitem[Girdhar et al.(2022)]{Girdhar.etal.2022} Girdhar, A., Harrison, C.~M., Mainieri, V., et al.\ 2022, \mnras, 512, 2, 1608

\bibitem[Goesaert et al.(2025)]{Goesaert.etal.2025} Goesaert, W.~M., Tristram, K.~R.~W., Impellizzeri, C.~M.~V., et al.\ 2025, \aap, 704, A125

\bibitem[Goold et al.(2024)]{Goold.etal.2024} Goold, K., Seth, A., Molina, M., et al.\ 2024, \apj, 966, 2, 204

\bibitem[Goold et al.(2026)]{Goold.etal.2026} Goold, K., Seth, A., Molina, M., et al.\ 2026, \apj, 1000, 2, 281

\bibitem[Guillard et al.(2009)]{Guillard.etal.2009} Guillard, P., Boulanger, F., Pineau Des For{\^e}ts, G., et al.\ 2009, \aap, 502, 2, 515

\bibitem[Guillard et al.(2012)]{Guillard.etal.2012} Guillard, P., Ogle, P.~M., Emonts, B.~H.~C., et al.\ 2012, \apj, 747, 95

\bibitem[Harrison \& Ramos Almeida(2024)]{Harrison&RamosAlmeida2024} Harrison, C.~M. \& Ramos Almeida, C.\ 2024, Galaxies, 12, 2, 17

\bibitem[Heckman \& Best(2014)]{Heckman&Best2014} Heckman, T.~M. \& Best, P.~N.\ 2014, \araa, 52, 589

\bibitem[Hermosa Mu{\~n}oz et al.(2026)]{Hermosa-Munoz.etal.2026} Hermosa Mu{\~n}oz, L., Gonz{\'a}lez Fern{\'a}ndez, J.~R., Alonso-Herrero, A., et al.\ 2026, \aap, 708, A297

\bibitem[Ho(2008)]{Ho2008} Ho, L.~C.\ 2008, \araa, 46, 475

\bibitem[Ho(2009)]{Ho2009} Ho, L.~C.\ 2009, \apj, 699, 1, 626

\bibitem[Ho et al.(2003)]{Ho.etal.2003} Ho, L.~C., Filippenko, A.~V., \& Sargent, W.~L.~W.\ 2003, \apj, 583, 1, 159

\bibitem[Ho et al.(2009)]{Ho.etal.2009} Ho, L.~C., Greene, J.~E., Filippenko, A.~V., et al.\ 2009, \apjs, 183, 1, 1

\bibitem[Hollenbach \& McKee(1989)]{Hollenbach&McKee1989} Hollenbach, D. \& McKee, C.~F.\ 1989, \apj, 342, 306

\bibitem[Holm et al.(2011)]{Holm.etal.2011} Holm, A.~I.~S., Johansson, H.~A.~B., Cederquist, H., et al.\ 2011, \jcp, 134, 4, 044301

\bibitem[Hopkins et al.(2008)]{Hopkins.etal.2008} Hopkins, P.~F., Hernquist, L., Cox, T.~J., et al.\ 2008, \apjs, 175, 2, 356

\bibitem[Hsieh et al.(2011)]{Hsieh.etal.2011} Hsieh, P.-Y., Matsushita, S., Liu, G., et al.\ 2011, \apj, 736, 2, 129

\bibitem[Hummel et al.(1987)]{Hummel.etal.1987} Hummel, E., van der Hulst, J.~M., \& Keel, W.~C.\ 1987, \aap, 172, 32

\bibitem[Izumi et al.(2018)]{Izumi.etal.2018} Izumi, T., Wada, K., Fukushige, R., et al.\ 2018, \apj, 867, 1, 48

\bibitem[Jensen et al.(2017)]{Jensen.etal.2017} Jensen, J.~J., H{\"o}nig, S.~F., Rakshit, S., et al.\ 2017, \mnras, 470, 3, 3071

\bibitem[Kennicutt et al.(2003)]{Kennicutt.etal.2003} Kennicutt, R.~C., Armus, L., Bendo, G., et al.\ 2003, \pasp, 115, 810, 928

\bibitem[Kristensen et al.(2023)]{Kristensen.etal.2023} Kristensen, L.~E., Godard, B., Guillard, P., et al.\ 2023, \aap, 675, A86

\bibitem[Koo et al.(2016)]{Koo.etal.2016} Koo, B.-C., Raymond, J.~C., \& Kim, H.-J.\ 2016, Journal of Korean Astronomical Society, 49, 3, 109

\bibitem[Kormendy \& Ho(2013)]{Kormendy&Ho2013} Kormendy, J. \& Ho, L.~C.\ 2013, \araa, 51, 1, 511

\bibitem[Lai et al.(2022)]{Lai.etal.2022} Lai, T.~S.-Y., Armus, L., U, V., et al.\ 2022, \apjl, 941, 2, L36

\bibitem[LaMassa et al.(2012)]{LaMassa.etal.2012} LaMassa, S.~M., Heckman, T.~M., Ptak, A., et al.\ 2012, \apj, 758, 1, 1

\bibitem[Law et al.(2023)]{Law.etal.2023} Law, D.~R., E. Morrison, J., Argyriou, I., et al.\ 2023, \aj, 166, 2, 45

\bibitem[Leroy et al.(2023)]{Leroy.etal.2023} Leroy, A.~K., Sandstrom, K., Rosolowsky, E., et al.\ 2023, \apjl, 944, 2, L9

\bibitem[Lofaro et al.(2026)]{Lofaro.etal.2026} Lofaro, C.~M., D{\'\i}az Santos, T., Shivaei, I., et al.\ 2026, arXiv:2606.18230

\bibitem[Li(2020)]{Li2020} Li, A.\ 2020, Nature Astronomy, 4, 339

\bibitem[L{\'o}pez et al.(2025)]{Lopez.etal.2025} L{\'o}pez, I.~E., Bertola, E., Reynaldi, V., et al.\ 2025, \aap, 704, A88

\bibitem[Lopez-Rodriguez et al.(2026)]{Lopez-Rodriguez.etal.2026} Lopez-Rodriguez, E., Sanchez-Bermudez, J., Gonz{\'a}lez-Mart{\'\i}n, O., et al.\ 2026, Nature Communications, 17, 1, 42

\bibitem[Maragkoudakis et al.(2026)]{Maragkoudakis.etal.2026} Maragkoudakis, A., Boersma, C., Peeters, E., et al.\ 2026, \aap, 709, A38

\bibitem[Maragkoudakis et al.(2025)]{Maragkoudakis.etal.2025} Maragkoudakis, A., Boersma, C., Temi, P., et al.\ 2025, \apj, 979, 1, 90

\bibitem[Maragkoudakis et al.(2018)]{Maragkoudakis.etal.2018} Maragkoudakis, A., Ivkovich, N., Peeters, E., et al.\ 2018, \mnras, 481, 5370

\bibitem[Marconi et al.(1994)]{Marconi.etal.1994} Marconi, A., Moorwood, A.~F.~M., Origlia, L., et al.\ 1994, The Messenger, 78, 20

\bibitem[Marshall et al.(2007)]{Marshall.etal.2007} Marshall, J.~A., Herter, T.~L., Armus, L., et al.\ 2007, \apj, 670, 1, 129

\bibitem[McNamara \& Nulsen(2007)]{McNamara&Nulsen2007} McNamara, B.~R. \& Nulsen, P.~E.~J.\ 2007, \araa, 45, 1, 117

\bibitem[Mezcua \& Prieto(2014)]{Mezcua&Prieto2014} Mezcua, M. \& Prieto, M.~A.\ 2014, \apj, 787, 1, 62

\bibitem[Micelotta et al.(2010a)]{Micelotta.etal.2010a} Micelotta, E.~R., Jones, A.~P., \& Tielens, A.~G.~G.~M.\ 2010a, \aap, 510, A36

\bibitem[Micelotta et al.(2010b)]{Micelotta.etal.2010b} Micelotta, E.~R., Jones, A.~P., \& Tielens, A.~G.~G.~M.\ 2010b, \aap, 510, A37

\bibitem[Morganti et al.(2013a)]{Morganti.etal.2013a} Morganti, R., Fogasy, J., Paragi, Z., et al.\ 2013a, Science, 341, 6150, 1082

\bibitem[Morganti et al.(2013b)]{Morganti.etal.2013b} Morganti, R., Frieswijk, W., Oonk, R.~J.~B., et al.\ 2013b, \aap, 552, L4

\bibitem[Mouri et al.(2000)]{Mouri.etal.2000} Mouri, H., Kawara, K., \& Taniguchi, Y.\ 2000, \apj, 528, 1, 186

\bibitem[Nagar et al.(2005)]{Nagar.etal.2005} Nagar, N.~M., Falcke, H., \& Wilson, A.~S.\ 2005, \aap, 435, 2, 521

\bibitem[Nemmen et al.(2014)]{Nemmen.etal.2014} Nemmen, R.~S., Storchi-Bergmann, T., \& Eracleous, M.\ 2014, \mnras, 438, 4, 2804

\bibitem[Nesvadba et al.(2017)]{Nesvadba.etal.2017} Nesvadba, N.~P.~H., Drouart, G., De Breuck, C., et al.\ 2017, \aap, 600, A121

\bibitem[O'Dowd et al.(2009)]{ODowd.etal.2009} O'Dowd, M.~J., Schiminovich, D., Johnson, B.~D., et al.\ 2009, \apj, 705, 885

\bibitem[Ogle et al.(2010)]{Ogle.etal.2010} Ogle, P., Boulanger, F., Guillard, P., et al.\ 2010, \apj, 724, 2, 1193

\bibitem[Ogle et al.(2025)]{Ogle.etal.2025} Ogle, P.~M., Sebastian, B., Aravindan, A., et al.\ 2025, \apj, 983, 2, 98

\bibitem[Pantoni et al.(2026)]{Pantoni.etal.2026} Pantoni, L., Baes, M., Decin, L., et al.\ 2026, arXiv:2603.23674

\bibitem[Pellegrini et al.(2002)]{Pellegrini.etal.2002} Pellegrini, S., Fabbiano, G., Fiore, F., et al.\ 2002, \aap, 383, 1

\bibitem[Pillepich et al.(2018)]{Pillepich.etal.2018} Pillepich, A., Springel, V., Nelson, D., et al.\ 2018, \mnras, 473, 3, 4077

\bibitem[Pontoppidan et al.(2024)]{Pontoppidan.etal.2024} Pontoppidan, K.~M., Salyk, C., Banzatti, A., et al.\ 2024, \apj, 963, 2, 158

\bibitem[Ramos Almeida et al.(2022)]{RamosAlmeida.etal.2022} Ramos Almeida, C., Bischetti, M., Garc{\'\i}a-Burillo, S., et al.\ 2022, \aap, 658, A155

\bibitem[Ramos Almeida et al.(2023)]{RamosAlmeida.etal.2023} Ramos Almeida, C., Esparza-Arredondo, D., Gonz{\'a}lez-Mart{\'\i}n, O., et al.\ 2023, \aap, 669, L5

\bibitem[Ramos Almeida et al.(2025)]{RamosAlmeida.etal.2025} Ramos Almeida, C., Garc{\'\i}a-Bernete, I., Pereira-Santaella, M., et al.\ 2025, \aap, 698, A194

\bibitem[Richings \& Faucher-Gigu{\`e}re(2018a)]{Richings&Faucher-Giguere2018a} Richings, A.~J. \& Faucher-Gigu{\`e}re, C.-A.\ 2018a, \mnras, 478, 3, 3100

\bibitem[Richings \& Faucher-Gigu{\`e}re(2018b)]{Richings&Faucher-Giguere2018b} Richings, A.~J. \& Faucher-Gigu{\`e}re, C.-A.\ 2018b, \mnras, 474, 3, 3673

\bibitem[Riffel et al.(2021)]{Riffel.etal.2021} Riffel, R.~A., Bianchin, M., Riffel, R., et al.\ 2021, \mnras, 503, 4, 5161

\bibitem[Riffel et al.(2026a)]{Riffel.etal.2026a} Riffel, R.~A., Colina, L., Costa-Souza, J.~H., et al.\ 2026a, \aap, 705, A59

\bibitem[Riffel et al.(2026b)]{Riffel.etal.2026b} Riffel, R.~A., Souza-Oliveira, G.~L., Colina, L., et al.\ 2026b, arXiv:2605.02663

\bibitem[Riffel et al.(2020)]{Riffel.etal.2020} Riffel, R.~A., Zakamska, N.~L., \& Riffel, R.\ 2020, \mnras, 491, 1, 1518

\bibitem[Rigopoulou et al.(2021)]{Rigopoulou.etal.2021} Rigopoulou, D., Barale, M., Clary, D.~C., et al.\ 2021, \mnras, 504, 5287

\bibitem[Rigopoulou et al.(2024)]{Rigopoulou.etal.2024} Rigopoulou, D., Donnan, F.~R., Garc{\'\i}a-Bernete, I., et al.\ 2024, \mnras, 532, 2, 1598

\bibitem[Roche et al.(2006)]{Roche.etal.2006} Roche, P.~F., Packham, C., Telesco, C.~M., et al.\ 2006, \mnras, 367, 4, 1689

\bibitem[Rodr{\'\i}guez-Ardila et al.(2017)]{Rodriguez-Ardila.etal.2017} Rodr{\'\i}guez-Ardila, A., Mason, R.~E., Martins, L., et al.\ 2017, \mnras, 465, 1, 906

\bibitem[Roussel et al.(2007)]{Roussel.etal.2007} Roussel, H., Helou, G., Hollenbach, D.~J., et al.\ 2007, \apj, 669, 959

\bibitem[Roy et al.(2026)]{Roy.etal.2026} Roy, N., Heckman, T., Henry, A., et al.\ 2026, \apj, 1002, 1, 57

\bibitem[Rupke \& Veilleux(2011)]{Rupke&Veilleux2011} Rupke, D.~S.~N. \& Veilleux, S.\ 2011, \apjl, 729, 2, L27

\bibitem[Sales et al.(2010)]{Sales.etal.2010} Sales, D.~A., Pastoriza, M.~G., \& Riffel, R.\ 2010, \apj, 725, 605

\bibitem[Seabold \& Perktold(2010)]{Seabold&Perktold2010} Seabold, S., \& Perktold, J. 2010, in Proc. of the 9th Python 558 in Science Conf., ed. S. van der Walt \& J. Millman (Austin, TX: Enthought, Inc.), 92

\bibitem[Shipley et al.(2016)]{Shipley.etal.2016} Shipley, H.~V., Papovich, C., Rieke, G.~H., et al.\ 2016, \apj, 818, 60

\bibitem[Silk \& Rees(1998)]{Silk&Rees1998} Silk, J. \& Rees, M.~J.\ 1998, \aap, 331, L1

\bibitem[Smith et al.(2007)]{Smith.etal.2007} Smith, J.~D.~T., Draine, B.~T., Dale, D.~A., et al.\ 2007, \apj, 656, 2, 770

\bibitem[Storchi-Bergmann et al.(2009)]{Storchi-Bergmann.etal.2009} Storchi-Bergmann, T., McGregor, P.~J., Riffel, R.~A., et al.\ 2009, \mnras, 394, 3, 1148

\bibitem[Tabatabaei et al.(2018)]{Tabatabaei.etal.2018} Tabatabaei, F.~S., Minguez, P., Prieto, M.~A., et al.\ 2018, Nature Astronomy, 2, 83

\bibitem[Thean et al.(2000)]{Thean.etal.2000} Thean, A., Pedlar, A., Kukula, M.~J., et al.\ 2000, \mnras, 314, 3, 573

\bibitem[Tielens(2008)]{Tielens2008} Tielens, A.~G.~G.~M.\ 2008, \araa, 46, 289

\bibitem[Togi \& Smith(2016)]{Togi&Smith2016} Togi, A. \& Smith, J.~D.~T.\ 2016, \apj, 830, 18

\bibitem[Treyer et al.(2010)]{Treyer.etal.2010} Treyer, M., Schiminovich, D., Johnson, B.~D., et al.\ 2010, \apj, 719, 1191

\bibitem[Van De Putte et al.(2025)]{VanDePutte.etal.2025} Van De Putte, D., Peeters, E., Gordon, K.~D., et al.\ 2025, \aap, 701, A111

\bibitem[Veilleux et al.(2005)]{Veilleux.etal.2005} Veilleux, S., Cecil, G., \& Bland-Hawthorn, J.\ 2005, \araa, 43, 1, 769

\bibitem[Venturi et al.(2021)]{Venturi.etal.2021} Venturi, G., Cresci, G., Marconi, A., et al.\ 2021, \aap, 648, A17

\bibitem[Venturi et al.(2023)]{Venturi.etal.2023} Venturi, G., Treister, E., Finlez, C., et al.\ 2023, \aap, 678, A127

\bibitem[Voit(1992)]{Voit1992} Voit, G.~M.\ 1992, \mnras, 258, 841

\bibitem[Weaver et al.(2010)]{Weaver.etal.2010} Weaver, K.~A., Mel{\'e}ndez, M., Mushotzky, R.~F., et al.\ 2010, \apj, 716, 2, 1151

\bibitem[Weinberger et al.(2017)]{Weinberger.etal.2017} Weinberger, R., Springel, V., Hernquist, L., et al.\ 2017, \mnras, 465, 3, 3291

\bibitem[Wells et al.(2015)]{Wells.etal.2015} Wells, M., Pel, J.-W., Glasse, A., et al.\ 2015, \pasp, 127, 953, 646

\bibitem[Wright et al.(2023)]{Wright.etal.2023} Wright, G.~S., Rieke, G.~H., Glasse, A., et al.\ 2023, \pasp, 135, 1046, 048003

\bibitem[Xie \& Ho(2019)]{Xie&Ho2019} Xie, Y. \& Ho, L.~C.\ 2019, \apj, 884, 136

\bibitem[Xie \& Ho(2022)]{Xie&Ho2022} Xie, Y. \& Ho, L.~C.\ 2022, \apj, 925, 218

\bibitem[Yuan \& Narayan(2014)]{Yuan&Narayan2014} Yuan, F. \& Narayan, R.\ 2014, \araa, 52, 529

\bibitem[Yui Dan et al.(2026)]{YuiDan.etal.2026} Yui Dan, K., Seebeck, J., Veilleux, S., et al.\ 2026, arXiv:2605.03016

\bibitem[Zhang \& Ho(2023a)]{Zhang&Ho2023a} Zhang, L. \& Ho, L.~C.\ 2023a, \apj, 943, 1, 1

\bibitem[Zhang \& Ho(2023b)]{Zhang&Ho2023b} Zhang, L. \& Ho, L.~C.\ 2023b, \apj, 943, 1, 60

\bibitem[Zhang \& Ho(2023c)]{Zhang&Ho2023c} Zhang, L. \& Ho, L.~C.\ 2023c, \apjl, 953, 1, L9

\bibitem[Zhang et al.(2025)]{Zhang.etal.2025} Zhang, L., Davies, R.~I., Packham, C., et al.\ 2025, \apjs, 280, 2, 65

\bibitem[Zhang et al.(2024b)]{Zhang.etal.2024b} Zhang, L., Garc{\'\i}a-Bernete, I., Packham, C., et al.\ 2024b, \apjl, 975, 1, L2

\bibitem[Zhang et al.(2022)]{Zhang.etal.2022} Zhang, L., Ho, L.~C., \& Li, A.\ 2022, \apj, 939, 22

\bibitem[Zhang et al.(2024a)]{Zhang.etal.2024a} Zhang, L., Packham, C., Hicks, E.~K.~S., et al.\ 2024a, \apj, 974, 2, 195

\bibitem[Zhang et al.(2026)]{Zhang.etal.2026} Zhang, L., Packham, C., Hicks, E.~K.~S., et al.\ 2026, \apjl, 998, 2, L32

\bibitem[Zhuang et al.(2019)]{Zhuang&Ho2019} Zhuang, M.-Y., Ho, L.~C., \& Shangguan, J.\ 2019, \apj, 873, 2, 103

\end{thebibliography}
\end{document}